\documentclass[12pt]{article}
\pdfoutput=1
\usepackage{amsmath,amsfonts,amssymb,amsthm,amssymb,bbm,bm}
\usepackage{pdfsync}
\usepackage{graphicx,import}
\usepackage[utf8]{inputenc}
\usepackage{empheq}
\usepackage[all]{xy}
\usepackage{stmaryrd}
\usepackage{rotating}
\usepackage[dvipsnames]{xcolor}  
\usepackage{slashed}
\usepackage{cancel}
\usepackage{array, makecell} %
\usepackage{comment}
\usepackage{tikz-cd}
\usepackage{hhline}
\usepackage{cancel}
\usepackage{dsfont}
\usepackage{stackengine}
\usepackage[breakable]{tcolorbox}
\usepackage{mathrsfs}
\usepackage{booktabs}
\usepackage{float}
\usepackage{changepage}
\usepackage[normalem]{ulem}

\allowdisplaybreaks

\usepackage[pagebackref]{hyperref}
\renewcommand*{\backref}[1]{}
\renewcommand*{\backrefalt}[4]{\small{
    \ifcase #1%
          \or (Cited on page~#2.)%
          \else (Cited on pages~#2.)%
    \fi%
    }}
\hypersetup{colorlinks,linkcolor={RedViolet},citecolor={Blue},urlcolor={Blue}}

\DeclareMathAlphabet\mathrsfso{U}{rsfso}{m}{n}

\usepackage[scr=boondoxo,scrscaled=1.1]{mathalpha}

\makeatletter

\DeclareFontFamily{U}{matha}{\hyphenchar\font45}
\DeclareFontShape{U}{matha}{m}{n}{
      <5> <6> <7> <8> <9> <10> gen * matha
      <10.95> matha10 <12> <14.4> <17.28> <20.74> <24.88> matha12
      }{}
\DeclareSymbolFont{matha}{U}{matha}{m}{n}
\DeclareMathSymbol{\oright}       {2}{matha}{"69}

\usepackage[top=2cm, bottom=2.5cm, left=2.5cm, right=2.5cm]{geometry}

\newcommand{\doublehat}[1]{%
\begingroup%
  \let\macc@kerna\z@%
  \let\macc@kernb\z@%
  \let\macc@nucleus\@empty%
  \hat{\raisebox{.55ex}{\vphantom{\ensuremath{#1}}}\smash{\hat{#1}}}%
\endgroup%
}

\renewcommand{\ni}{\noindent}

\newcommand{\bit}{\begin{itemize}}
\newcommand{\eit}{\end{itemize}}
\newcommand{\bd}{\begin{description}}
\newcommand{\ed}{\end{description}}

\newcommand{\bc}{\begin{center}}
\newcommand{\ec}{\end{center}}

\newcommand{\C}{{\mathbb C}}
\newcommand{\N}{{\mathbb N}}

\newcommand{\gbms}{\mathfrak{gbms}}
\newcommand{\ebms}{\mathfrak{ebms}}

\newcommand{\hG}{\hat{G}}

\def\be#1\ee{\begin{align}#1\end{align}}
\newcommand{\bea}{\begin{eqnarray}}
\newcommand{\eea}{\end{eqnarray}}
\newcommand{\bs}{\begin{subequations}}
\newcommand{\es}{\end{subequations}}
\newcommand{\nn}{\nonumber}

\newcommand{\sds}{\,\tikz[baseline=1]{
		\draw[line width=.6pt] (0,.13) circle (.8ex);
		\draw[line width=.6pt] (-.05ex,.27) -- (-.05ex,0);
		\draw[line width=.6pt] (0,.13) -- (.8ex,.13);}\,}

\usepackage[numbers,sort&compress]{natbib}

\def\rd{\mathrm{d}}
\def\pa{\partial}
\newcommand{\Dcal}{\mathcal{D}}

\newcommand{\dT}{\delta_T}
\newcommand{\dTp}{\delta_{T'}}

\newcommand{\hdT}{\hat{\delta}_T}
\newcommand{\hdTp}{\hat{\delta}_{T'}}

\newcommand{\LL}{\mathscr{L}}

\newcommand{\overbar}[1]{\mkern 3mu\overline{\mkern-3.5mu#1\mkern-2mu}\mkern 2mu}

\newcommand{\ooverline}[1]{\mkern 3mu\overline{\overline{#1}}\mkern 2mu}

\newcommand{\bC}{\overbar{C}}

\newcommand{\bz}{\bar{z}}
\renewcommand{\bm}{\overbar{m}}
\newcommand{\bY}{\overbar{Y}}
\newcommand{\bD}{\overbar{D}}
\newcommand{\bP}{\overbar{P}}
\newcommand{\Vs}{\overline{\mathsf{V}}(S)}
\newcommand{\Vss}{\ooverline{\mathsf{V}}(S)}
\newcommand{\Ws}{\overline{\mathsf{W}}(S)}
\newcommand{\Wo}{\overline{\mathsf{W}}}

\newcommand{\bWcal}{\overline{\Wcal}}

\newcommand{\bbVcal}{\ooverline{\Vcal}}

\newcommand{\bJ}{\bar{J}}

\newcommand{\hV}{\widehat{\V}}

\newcommand{\tT}{\widetilde{T}}

\newcommand{\X}{\mathfrak{X}}

\newcommand{\sn}{\mathsf{SN}}

\newcommand{\V}{\mathsf{V}}

\newcommand{\Vcal}{\mathcal{V}}

\newcommand{\W}{\mathsf{W}}
\newcommand{\Wcal}{\mathcal{W}}

\newcommand{\Ccel}[1]{\mathcal{C}^\textsf{cel}_{#1}(S)}

\renewcommand{\l}{\ell}
\newcommand{\sshp}{Y^s_{\l,m}}

\newcommand{\ad}{{\textrm{ad}}_\sigma}

\newcommand{\adtG}{{\textrm{ad}}_{\sigma_t} \hat{G}_t}
\newcommand{\adaG}{{\textrm{ad}}_{\sigma_\alpha} \hat{G}_\alpha}
\newcommand{\ada}{\textrm{ad}_{\sigma_\alpha}}

\newcommand{\mTG}{\mathcal{T}^G}

\newcommand{\mTaG}{\mathcal{T}^{G_\alpha}}

\renewcommand{\t}[1]{\tau_{#1}}
\newcommand{\T}[1]{T_{#1}}

\newcommand{\Tp}[1]{T'_{#1}}

\newcommand{\Tpp}[1]{T''_{#1}}

\newcommand{\scri}{\mathrsfso{I}}

\newcommand{\poisson}[1]{\big\{#1\big\}}
\newcommand{\paren}[1]{\,\Lbag #1\Rbag\,}

\newcommand{\cyc}{\overset{\circlearrowleft}{=}}

\newcommand{\Ham}{\eta}

\newcommand{\hHam}{\eta_{\scriptscriptstyle{G}}}
\newcommand{\hHama}{\eta_{\scriptscriptstyle{G}_\alpha}}

\newcommand{\bfm}[1]{\boldsymbol{#1}}

\usepackage{tikz}
\newcommand*\circled[1]{\tikz[baseline=(char.base)]{
            \node[shape=circle,draw,inner sep=2pt] (char) {#1};}}
       
\definecolor{myorange}{RGB}{223, 109, 20}

\definecolor{argile}{RGB}{239, 239, 239}
\definecolor{beige}{RGB}{254, 253, 240}

\begin{document}

\title{\Large{\bf Asymptotic Higher Spin Symmetries I:\\
Covariant Wedge Algebra in Gravity}}

\author{Nicolas Cresto$^{1,2}$\thanks{ncresto@perimeterinstitute.ca},
Laurent Freidel$^1$\thanks{lfreidel@perimeterinstitute.ca} 
}
\date{\small{\textit{
$^1$Perimeter Institute for Theoretical Physics,\\ 31 Caroline Street North, Waterloo, Ontario, N2L 2Y5, Canada\\ \smallskip
$^2$Department of Physics \& Astronomy, University of Waterloo,\\Waterloo, Ontario, N2L 3G1, Canada
}}}

\maketitle
\begin{abstract}
In this paper, we study gravitational symmetry algebras that live on 2-dimensional cuts $S$ of asymptotic infinity. 
We define a notion of wedge algebra $\Wcal(S)$ which depends on the topology of $S$.  
For the cylinder $S=\C^*$ we recover the celebrated $Lw_{1+\infty}$ algebra. 
For the 2-sphere $S^2$, the wedge algebra reduces to a central extension of the anti-self-dual projection of the Poincaré algebra.
We then extend $\Wcal(S)$ outside of the wedge space and build a new Lie algebra $\Wcal_\sigma(S)$, which can be viewed as a deformation of the wedge algebra by a spin two field $\sigma$ playing the role of the shear at a cut of $\scri$. 
This algebra represents the gravitational symmetry algebra in the presence of a non trivial shear and is characterized by a covariantized version of the wedge condition.
Finally, we construct a dressing map that provides a Lie algebra isomorphism between the covariant and regular wedge algebras. \\

\ni \textbf{Keywords:}
Gravity, Asymptotic Symmetries, Asymptotically Flat Spacetimes, Celestial Holography, Wedge Algebra.
\end{abstract}

\newpage
\tableofcontents
\newpage

\section{Introduction}

In the quest for understanding the symmetries of asymptotically flat spacetimes, the study of conformally soft gravitons symmetries led to the discovery of a $Lw_{1+\infty}$ algebra structure that organizes the OPEs of the soft modes \cite{Guevara:2021abz, Strominger:2021mtt, Himwich:2021dau, Ball:2021tmb}.
Understanding this algebraic structure is particularly relevant for the study of the gravitational $\mathcal{S}$-matrix, infrared physics and the relationship with asymptotic symmetries.

Historically, $W_N$ algebras have been studied in the context of two-dimensional conformal field theories as extensions of the Virasoro algebra to higher conformal spin operators  \cite{zamolodchikov1995infinite,PseudoDiffBakas}.
The contraction (or classical limit) of $W_\infty$ to $w_{\infty}$ \cite{BAKAS198957}, together with the enlargement to $w_{1+\infty}$ can be interpreted in terms of area preserving diffeomorphisms \cite{Boyer_1985, Shen:1992dd}.

The loop extension of $w_{1+\infty}$ denoted $Lw_{1+\infty}$ appears naturally in the twistorial approach of gravity as canonical transformations of the 2-dimensional fibers of the twistor fibration over the celestial sphere \cite{Dunajski:2000iq, Adamo:2021lrv, Mason:2022hly}. 
In the original Penrose's construction, self-dual spacetimes \cite{Penrose:1976js} are recast in terms of complex deformations of twistor space. 
Moreover, these deformations were obtained by applying a canonical transformation of the fiber on the overlap between the two coordinate patches. 
The fact that the topology of the base is $\C^*$ leads to a parametrization of the twistor deformations in terms of $Lw_{1+\infty}$. 
In this perspective, the latter algebra therefore parametrizes self-dual spacetimes. 
Recently, the connection between twistor and celestial OPE was achieved through the study of a twistor sigma model for self-dual gravity developed in \cite{Adamo:2021bej,Adamo:2021lrv, Mason:2022hly}.
More recent developments involving $Lw_{1+\infty}$ on the twistor side involve Moyal deformations \cite{Bu:2022iak}, the connection with the Carrollian perspective \cite{Mason:2023mti} and,  quite remarkably, the relevance of $Lw_{1+\infty}$ as a symmetry algebra for de-Sitter amplitudes 
\cite{Taylor:2023ajd, Bittleston:2024rqe}.

 Another point of view on $Lw_{1+\infty}$ follows the more traditional developments of asymptotic symmetries and the understanding of soft theorems as conservation laws \cite{Bondi:1962px,Sachs:1962wk,Strominger:2017zoo, Raclariu:2021zjz}. In this approach, one identifies a collection of higher spin charges aspects that generalize the Bondi mass aspect (spin $0$) and the angular momenta aspect (spin $1$) to higher spins \cite{Hamada:2018vrw, Freidel:2021dfs, Freidel:2021ytz, Geiller:2024bgf}. 
 The soft theorems are then understood as a matching condition for this set of higher spin charges at spacelike infinity. 
 
This point of view is closer to a Noetherian perspective where the charges aspects  are  symmetry generators which act on the gravitational phase space and are dual to a symmetry algebra.
The algebra that arises from this analysis is not directly $Lw_{1+\infty}$: it is a shifted version of the Schouten-Nijenhuis algebra for \textit{symmetric} contravariant tensor fields \cite{Schouten, SNbracket}.
The advantage of this description is that the symmetry generators do not have to be holomorphic or to have specific singularity properties on the complex plane, as is the case for $Lw_{1+\infty}$.
In fact, one can define the symmetry algebra on any Riemann surface $S$.
This algebra is encoded into a set of spin $s$ fields on $S$ denoted $T_s(z,\bz)$ for $s\geq -1$.  
$\T0$ and $\T1$ correspond respectively to the super-translations and super-Lorentz-rotations parameters.  
The parameters $T_{-1}$ labels the center.
The shifted Schouten-Nijenhuis algebra bracket is given by 
 \be 
 [T,T']^\V_s= \sum_{n=0}^{s+1} (n+1)\big(\T{n} D \Tp{s+1-n}-\Tp{n} D \T{s+1-n}\big),
 \ee 
where $D$ is the holomorphic covariant derivative on $S$.
If we truncate this algebra to spin $0$ and spin $1$ fields we recover the  $\gbms$ algebra which includes 2d diffeomorphisms  \cite{Barnich:2016lyg, Campiglia:2014yka, Compere:2018ylh, Campiglia:2020qvc, Freidel:2021fxf}.

In the current work, we establish that the consistent inclusion of the supertranslations into the \textit{higher spin}---i.e. not just super-rotations---algebra forces the parameters to satisfy the condition $D^{s+2} T_s=0$, which defines the \emph{wedge} sub-algebra denoted $\Wcal(S)$. 
As we will see, this condition is \emph{necessary} in order for the Jacobi identity to be valid.
The connection between $\Wcal(S)$ and $Lw_{1+\infty}$ is revealed if one chooses $S$ to be the cylinder. 
Indeed, it was established in \cite{Donnay:2024qwq} that $\Wcal(\C^*) = Lw_{1+\infty}$. 
The correspondence between the twistor and celestial sphere actions is such that the wedge condition on $S=\C^*$ follows from the demand for holomorphicity of the twistor transformation parameters.
On the other hand, if we choose $S=S^2$ as the global 2-sphere then we find that $\Wcal(S^2)$ is a finite-dimensional algebra given by a 3-dimensional central extension of the chiral Poincaré algebra $\mathfrak{sl}_-(2,\C) \sds \C^4$ (see section \ref{sec:Wspace} and \ref{Sec:Walgebra}), a fact first noted in \cite{Krasnov:2021cva}.
This means that the wedge algebra is a global symmetry algebra that depends on the 2d surface topology.

Moreover, the higher spin fields $s\geq 1$ form a Lie sub-algebra that extends $\mathrm{Diff}(S)$ to higher spins.
The wedge condition is \textit{not} required in this case.
This explains why \cite{Freidel:2023gue} had been able to relax the wedge restriction at spin 2 when checking the closure of the algebra at quadratic order.
This property thus clarifies why their computation is not in contradiction with the aforementioned Jacobi identity anomaly.
The latter vanishes in the absence of time (namely spin 0 parameter).

Note that the direct proof of the equivalence between the twistorial and canonical realization of the $Lw_{1+\infty}$ symmetry was first given in \cite{Donnay:2024qwq} at the quadratic order in $G_{\! N}$. 
Recently, in a beautiful work, this connection was established further to all order in \cite{Kmec:2024nmu}.

In the present manuscript, we investigate the possibility of deforming the wedge algebra.
There are several reasons to expect such a deformation. The main one is physical: While the wedge algebra is a perturbative symmetry of amplitudes near the flat space vacuum, we expect the symmetry to exist at any non-radiative cut of $\scri$. Due to the memory effect, we know that such cuts will carry a non-trivial value of the shear $\sigma$. Therefore, promoting the wedge algebra as a symmetry algebra of gravity requires us to find what symmetry algebra is represented at cuts with non-zero shear.

Another reason comes from the analysis done in \cite{Freidel:2023gue}, which suggested that it was possible to have a canonical representation of the symmetry algebra beyond the wedge. 
This suggests that a deformation of the wedge algebra beyond the wedge should exist. 
The challenge is that the wedge condition appears as a necessary condition for the Jacobi identity.

Quite remarkably, a deformation of the wedge algebra exists, which respects Jacobi and is labeled by a shear parameter. The bracket for this algebra, denoted $\Wcal_\sigma(S)$, reads 
\begin{equation}
    [T,T']^\sigma_s = 
    [T,T']^\V_s-(s+3)\sigma\big(\T0\Tp{s+2}-\Tp0\T{s+2}\big). \label{sigmaBracketIntro}
\end{equation}
This bracket respects the Jacobi identity for a set of parameters that satisfy a deformed wedge condition (see sec. \ref{sec:NewFieldDepBracket}). 
In our final result, we show that this deformed algebra is equivalent to the undeformed one through a \emph{dressing} transformation, which allows us to describe symmetry parameters in $\Wcal_\sigma(S)$ in terms of dressed symmetry parameters in $\Wcal(S)$. The possibility of removing the shear through a redefinition of the symmetry parameter provides an algebraic equivalent of Newman's $\mathcal{H}$-theorem and good cut equation (see \cite{newman_heaven_1976, Adamo:2009vu}), which established that we can reabsorb the asymptotic shear into a redefinition of the cut. 

One of the primary motivations for the work presented here is the possibility to show that the dressed algebra $\Wcal_\sigma(S)$ can be canonically and non-linearly represented on the gravitational phase space. This result is  shown in our companion paper \cite{Cresto:2024mne} and relates to the recent twistorial canonical analysis done in \cite{Kmec:2024nmu}.

The paper is organized as follows:
In section \ref{sec:Preliminaries}, we introduce our conventions of notation and give a reminder about conformal and spin weights. 
We then propose in section \ref{sec:WedgeAlgebra} a generalization of the definition of the $w_{1+\infty}$ bracket, valid for any two-dimensional manifold with arbitrary genus.
We refer to this bracket as the $\W$-bracket, associated to the Lie algebra $\Wcal(S)$.
When $S\equiv\C^*$, we recover the wedge $Lw_{1+\infty}$ algebra, while onto the sphere, we get the central extension of the anti-self-dual projection of the Poincaré algebra. 
We emphasize the role of the degree $-1$ elements which are central elements.  
In section  \ref{sec:DeformingWedge} we introduce the deformation of the wedge algebra by the shear $\sigma(z,\bz)$. 
This comes after a discussion about Jacobi identity violation for parameters outside of the wedge space.
We also introduce the notion of a  shear dependent covariant derivative that represents the adjoint action of the time translation elements. 
Besides, we use this covariant derivative to present equivalent forms of the $\sigma$-bracket, which highlight some of its key properties.
To finish, in section \ref{sec:CovWedgeSol}, we prove the isomorphism between $\Wcal(S)$ and $\Wcal_\sigma(S)$  which involves the definition of a dressing map $\mTG$ as a path ordered exponential along a flow in the space of Lie algebras, where the flow of the $\sigma$-bracket is determined by the adjoint action of the Goldstone field $G$.

\section{Preliminaries \label{sec:Preliminaries}}

\subsection{Notation}

We use the usual Bondi coordinates $(u,r,z,\bz)$ together with the null dyad $(m,\bm)$ on the sphere. 
We choose the normalization as follows:\footnote{For instance, on a round sphere $m=\frac{1+z\bz}{\sqrt2}\pa_z\equiv P\pa_z$.} 
\begin{equation}
    m=P\pa_z,\qquad\bm=\bP\pa_{\bz},\qquad m^A\bm^B\gamma_{AB}=1, \label{mdef}
\end{equation}
for a prefactor $P=P(z,\bz)$ and where $\gamma_{AB}=m_A\bm_B+\bm_A m_B$ is the asymptotic sphere metric.
The measure on the sphere takes the form
\begin{equation}
\bfm{\epsilon}_S=\frac12\epsilon_{AB}\, \rd\sigma^A\wedge\rd\sigma^B=i\sqrt{\gamma}\,\rd z\wedge\rd\bz\qquad\mathrm{with}\qquad\epsilon_{AB}=i(\bm_A m_B-m_A\bm_B).
\end{equation}

Rather than using the covariant derivative on the sphere $D_A$ with a free index $A$, one can use projected quantities and work with spin-weighted operators. Namely, an operator $O_s$ of spin-weight, or helicity, $s$ is defined by
\begin{equation}
O_s\equiv O_{\left\langle A_1\ldots A_s\right\rangle}m^{A_1}\ldots m^{A_s}.
\end{equation}
Similarly, the operator $O_{-s}$ is defined by the same contraction with $m\rightarrow\bm$. Angle brackets stand for symmetric and traceless. The tracelessness condition allows us to map any symmetric traceless tensor to a Lorentz scalar of helicity $s$ and a Lorentz scalar of helicity $-s$. To make things more explicit, one can write\footnote{The isomorphism is justified by a counting argument: A symmetric tensor with $s$ indices in dimension 2 has $\binom{2+s-1}{s}=s+1$ independent components. 
Imposing it to be traceless adds $\binom{s-1}{s-2}=s-1$ constraints, so that the number of independent components reduces to two, encoded in the two helicities.}
\begin{equation}
O_{\left\langle A_1\ldots A_s\right\rangle}=O_s\,\bm_{A_1}\ldots\bm_{A_s}+O_{-s}m_{A_1}\ldots m_{A_s}.
\end{equation}
For instance, $C\equiv C_{AB}m^Am^B$ and its complex conjugate $\bC$ will denote the asymptotic shear. Similarly, the news tensor will be denoted by $N=\pa_u C$

We introduce the intrinsic-Cartan derivatives on the sphere\footnote{Historically the \textit{edth} differential operator has been defined with a different normalization \cite{ethop}.} $\sqrt{2}D=\eth$ and $\sqrt{2}\bD=\bar{\eth}$ defined as the sphere covariant derivative projected along the dyad. For instance, on an operator of spin-weight $s$,
\begin{equation}
DO_s\equiv m^{A_1}\ldots m^{A_s}m^A D_AO_{A_1\ldots A_s}.
\end{equation}
Changing $m^A\rightarrow\bm^A$, we get a similar definition for $\bD$. 
Using \eqref{mdef}, we have equivalently
\begin{subequations}
\label{intrinsic1}
\begin{align}
DO_s&=P\bP^{-s}\partial_z\big(\bP^{s}O_s\big),\\
\bD O_s&=\bP P^s\partial_{\bz}\big(P^{-s}O_s\big).
\end{align}
\end{subequations}

Like a function on the sphere which admits an expansion in terms of spherical harmonics $Y^0_{\l,m}$, a spin-weighed function of helicity $s$ admits an expansion in terms of spin-spherical harmonics $\sshp$ \cite{ethop}, where $\ell\geq |s|$,
\begin{equation}
    O_s=\sum_{\l=|s|}^\infty\sum_{m=-\l}^\l O_s^{\l,m}\sshp,\qquad O_s^{\l,m}\in\C. \label{sshExpansion}
\end{equation}
$D$ and $\bD$ act as raising and lowering operators for the helicity. 
Their action on the basis element $\sshp$ is given by
\begin{align} 
    D\sshp &=\sqrt{\frac{(\l-s)(\l+s+1)}{2}}Y^{s+1}_{\l,m}, \label{raisingD} \\ 
    \bD \sshp &=-\sqrt{\frac{(\l+s)(\l-s+1)}{2}}Y^{s-1}_{\l,m}, \label{loweringbD}
\end{align}
from which we also get 
\be 
\bD D \sshp=-\frac{(\l-s)(\l+s+1)}{2}\sshp,
\qquad 
D \bD \sshp= -\frac{(\l+s)(\l-s+1)}{2}\sshp. \label{casimirDD}
\ee

\subsection{Definition of conformal weight and helicity \label{Sec:Celest}}

In this paper we define $S_0$ to denote the 2d-sphere and $S_n$ to be the 2d-sphere minus $n$-points.
In the following we denote $S$ any one of the $S_n$ and we define $\mathcal{C}(S)$ to be the space of smooth functions on $S$.
The most studied cases are $S_0$ in the gravitational literature and the cylinder $S_2$ in the celestial and twistor literature.

We start by defining the notion of celestial primary fields $\Ccel{(\Delta,s)}$ on the 2-sphere (or the punctured 2-sphere, on which $D\to\pa_z$), where $\Delta$ is the celestial conformal dimension and $s$ the helicity (or spin-weight) as before.
By definition a $\phi(z,\bz) \in \Ccel{(\Delta,s)}$ transforms under sphere diffeomorphisms $\mathcal Y\in \mathrm{Diff}(S)$ as
\be 
\delta_{\mathcal Y}  \phi
&=\big(YD+\tfrac12(\Delta + s) DY\big)\phi+\big(\bY\bD + \tfrac12(\Delta - s)\bD\bY\big) \phi. \label{sDeltaDef}
\ee
This reduces to the usual definition of a conformal primary field in 2d \cite{francesco2012conformal, Barnich:2021dta, Donnay:2021wrk} when $Y$ ($\bY$) is (anti-)holomorphic.
Such fields commonly appear in the celestial holography context when transforming momentum space scattering amplitudes to the conformal (boost) basis by a Mellin transform, see \cite{Raclariu:2021zjz} and references therein.
Here we extend the definition of conformal primary fields to arbitrary sphere diffeomorphisms \cite{Freidel:2021dfs}.
What we call celestial fields also corresponds to the restriction of Carrollian fields at the cut $u=0$ of null infinity, with $u$ the Bondi time \cite{Freidel:2021dfs, Cresto:2024mne}.

$Y$ and $\bY$ are the projections of $\mathcal Y$ along $\bm$ and $m$ respectively.
When writing $\mathcal Y\in\mathrm{Diff}(S)$, we mean the full vector on the sphere, namely
\begin{equation}
    \mathcal Y=\mathcal Y^A\pa_A=(m^A\mathcal Y_A)\bD+(\bm^A\mathcal Y_A)D\equiv \bY\bD+YD.
\end{equation}
Notice that  the parameter $Y$ is the component of a holomorphic vector field $Y\pa_z\in \Gamma(T^{1,0}S)$ and comes with a negative helicity degree, i.e.  $Y$ is the contraction of $\mathcal Y$ with $\bm$.
Hence $Y$ has weight $(-1,-1)$\footnote{This is clear from the fact that \eqref{sDeltaDef}  matches with the Lie derivative $\LL$ when acting on a vector field. Indeed, if we take $\Wcal\in \mathrm{Diff}(S)$, then 
\begin{equation}
    \delta_{\Wcal}Y= \bm_C\big(\LL_{\mathcal W}Y\big)^C=\bm_C\big(\Wcal^AD_AY^C-Y^A D_AW^C\big)=\big(WD+\overbar W\bD\big)Y-YDW.
\end{equation}} while $\bY$ has weight $(-1,1)$.\footnote{Remark that in general the multivector $\t{s}=\tau^{\langle A_1\cdots A_s\rangle}\bm_{A_1}\cdots\bm_{A_s}$, which shall play the role of dual variable to the higher-spin charges, has spin $-s$.}

We can restrict the conformal transformation labeled by the vector $\mathcal Y$ to preserve the round sphere metric. 
$Y$ is then holomorphic, i.e. $\bD Y=0$.
In this case, the operators $D$ and $\bD$ are respectively operators of celestial weight $(1,1)$ and $(1,-1)$.\footnote{To extend this property to Diff$(S)$ we need to introduce super-rotation Goldstone field
$\Phi$ and a boost connection ${\mathcal D}O_{(\Delta,s)} 
= (D+(s+\Delta)  \Phi)O_{(\Delta,s)}$,
see \cite{Campiglia:2020qvc, Donnay:2021wrk, Freidel:2021dfs}.
}

Once a conformal structure is chosen one  introduces homogeneous coordinates $\lambda_\alpha$ on the sphere such that  $z= \lambda_1/\lambda_0$. 
One can understand the conformal primaries as homogeneous section $O(n,\bar{n})$ over $\mathbb{CP}_1$ \cite{Eastwood_Tod_1982, gelfand1966generalized}, where the homogeneity degrees are given by the celestial weights as $n=-(\Delta+s)$ and $\bar{n}= -(\Delta -s) $.   This means that 
\begin{equation}
    \phi(a \lambda_\alpha,\bar{a} \bar{\lambda}_{\dot{\alpha}})=a^{-(\Delta+s)}\bar a^{-(\Delta-s)}\phi(\lambda_\alpha, \bar{\lambda}_{\dot{\alpha}}),\qquad a\in\C^*.
\end{equation}
In this formulation the SL$(2,\mathbb{C})$ action on the sphere is linear on $\lambda_\alpha$.

\section{The wedge algebra \label{sec:WedgeAlgebra}}

There is by now ample evidence that the infinite dimensional Lie algebra $Lw_{1+\infty}$ is a symmetry algebra of perturbative gravity in the self dual sector \cite{Adamo:2021lrv,Donnay:2024qwq}. 
One direct way to see this is to use the twistor representation of self dual solutions and use the fact that $Lw_{1+\infty}= L(\mathrm{Ham}(\mathbb{C}^2))$ is a loop algebra over Hamiltonian vector field acting on the fiber of the twistor fibration $\mathbb{T}^* \to \mathbb{CP}_1$.

Another way to see this is to show that this algebra possesses a canonical action on the asymptotic data at $\scri$ \cite{Freidel:2021dfs, Freidel:2021ytz, Geiller:2024bgf, Freidel:2023gue}.

In this section, we remind the reader about the \textit{wedge} algebra, first studied in the twistor context by Penrose \cite{Penrose:1976js}, and in celestial holography context by Strominger and al.~\cite{Himwich:2021dau,Strominger:2021mtt,Guevara:2021abz} and then in the canonical framework by Freidel, Pranzetti and Raclariu \cite{Freidel:2021dfs,Freidel:2021ytz,Freidel:2023gue}. 
The connection between the twistor and canonical action of $Lw_{1+\infty}$ has been studied in \cite{Donnay:2024qwq, Kmec:2024nmu}.

\subsection{The $\V$-space \label{sec:Vspace}} 

We introduce the following space of  celestial fields:
\be 
\V_s\equiv  \Ccel{(-1,-s)}, \qquad 
\V(S)\equiv \bigoplus_{s=-1}^\infty\V_s.
\ee 
$V(S)$ is a graded vector space where the spin  $s$ denotes the grade. Note that $\V_s\equiv \V_s(S)$ depends on the topology of $S$. 
To avoid notational cluttering we keep this dependence implicit in the notation $\V_s$. 
The gradation of $\V(S)$ induces a filtration denoted 
\be \label{filtration}
\V^s(S) := \bigoplus_{n=-1}^s \V_n(S).
\ee

In the following we denote by $T$ the elements of $\V(S)$ and by $T_{s}$ its grade $s$ element.
In other words, $T$ corresponds to the series of spin weighted field  $(\T{s})_{s+1\in\N}$ and $\T{s}$ is the result of the evaluation map $\pi_s$ to the subspace of degree $s$:\footnote{The associated \textit{projection} operator is $\iota\circ\pi_s$.}
\begin{align}
    \pi_s :\,&\V(S)\to\V_s \nn\\
         &T\mapsto \T{s}.
\end{align}
Conversely, the inclusion map $\iota$,
\begin{align}
    \iota :\,&\V_s\to\V(S) \nn\\
         &\T{s}\mapsto T=\iota(\T{s}), \label{inclusionMap}
\end{align}
takes the element $\T{s}$ and sends it to the series $\iota(\T{s})=(0,\ldots,0,\T{s},0,\ldots)$ for which $\pi_p(T)=0$ for $p\neq s$ and $\pi_s(T)=\T{s}$.
Finally, for later convenience, we also define 
\begin{equation}
     \Vss:=\V(S)\backslash \{\V_{-1}\oplus\V_0\}=\bigoplus_{s=1}^\infty\V_s. \label{VstarstarDef}
\end{equation}

\subsection{The $\W$-space \label{sec:Wspace}}

An important subspace of $\V(S)$ corresponds to the \textit{wedge} vector space:
\begin{equation}
    \boxed{~\W(S):= \Big\{T\in \V(S)\,\big|\, D^{s+2}\T{s}=0,~s\geqslant -1\Big\} \subset \V(S)~}. \label{DefWedge}
\end{equation}
The $\W$-space inherits the grading of $\V(S)$, namely
\be 
\W(S)= \bigoplus_{s=-1}^\infty \W_s(S),
\ee 
where $\W_s(S):= \{ T_s\in \V_s\,|\, D^{s+2} T_s=0\}$.

Depending if we work on the sphere $S_0$, the complex plane $\C\equiv S_1$ or the cylinder $\C^*\equiv\C\backslash \{0\}\equiv S_2$,\footnote{Recall that $S_n$ generically denotes a 2-sphere with $n$ points removed.} the restriction $D^{s+2}\T{s}=0,~s\geqslant -1$, on the parameters $\T{s}$ takes different forms (see \cite{Barnich:2021dta} for a similar discussion about $\mathfrak{bms}$).

On the sphere the solution depends on the spin $s$. 
It is direct to establish that the space of solutions of $D^2 T_0=0$ is 4-dimensional,\footnote{All the vector spaces discussed in this section are complex vector spaces, over $\mathbb{C}.$} parametrized by a 4-vector $x^a\in\C^4$ and given by
\be \label{T0T}
T_0(x) = x_a  q^a (z,\bz), \qquad 
q(z,\bz) = \frac{1}{(1+|z|^2)} \big( 1+|z|^2, z+\bz, -i(z-\bz), 1-|z|^2\big).
\ee 
The vector $q$ is null and of the form $(1, n^i)$ where $n$ is normalized, $n^i n^j\delta_{ij}=1$, and represents a point on the sphere.  
The space of solutions of $D T_{-1}=0$ and $D^3 T_{1}=0$ are three  dimensional and respectively given by 
\be 
T_{-1}(y) = y^iD n_i, \qquad T_{1}(w)= w^i\bD n_i, \qquad y,w\in\C^3.\label{TDT}
\ee 
On the other hand, the space of solutions  of $D^{s+2} T_{s}=0$ for $s\geq 2$ is zero-dimensional. It only  contains the null solution $T_s=0$.
The reader can check the appendix \ref{AppWedgeOnSphere} for a  proof using spherical harmonics.

This result means that on the sphere the wedge algebra is 10 dimensional. 
We will see around equation \eqref{central-1} that the algebra reduces to a central extension of the  chiral Poincaré group. The chiral Poincaré algebra is obtained if we restrict to $s\geq 0$.

On the complex plane $\C\equiv S_1$, the situation is different.
The wedge condition $D^{s+2}\T{s}=0$ is merely the anti-holomorphicity of $D^{s+1}\T{s}$ (since $D\to\pa_z$).
$\T{s}$ is thus an arbitrary entire series in $\bz$ and a polynomial of degree $s+1$ in $z$:
\begin{equation}    
\T{s}=\sum_{m=0}^{s+1} T_{(s,m)}(\bz) z^m, \qquad 
T_{(s,m)}(\bz) = \sum_{k=0}^\infty T_{(s,m,k)} \,\bz^k.
\end{equation}
The algebra it generates is $L_+(w_{1+\infty})$, a sub-algebra of $Lw_{1+\infty}$.

Finally, on $S_2\equiv \C^*$, we are allowed to take a Laurent series in $\bz$.\footnote{Recall that since $D(1/\bz)=2\pi\delta^{(2)}(z)$, we had to discard these solutions on $S_1$.}
Therefore, the general solution of $D^{s+2}\T{s}=0$ is
\begin{equation}    
\T{s}=\sum_{m=0}^{s+1} T_{(s,m)}(\bz) z^m, \qquad 
T_{(s,m)}(\bz) = \sum_{k=-\infty}^\infty T_{(s,m,k)} \,\bz^k. \label{TsOnS2}
\end{equation}
This generates $Lw_{1+\infty}$.
Notice that while the vector space $W(S_0)$ is finite dimensional, $W(S_1)$ and $W(S_2)$ are infinite dimensional.

The relationship between  $T_s$ and the modes commonly used in the celestial literature (see e.g. \cite{Strominger:2021mtt}) arises if we take $s=2p-3$ and write \eqref{TsOnS2} as
\begin{equation}
    \T{s}= \sum_{m=1-p}^{p-1}\sum_{k=-\infty}^\infty T_{(2p-3,p-1+m,k)}  z^{p-1+m}\bz^k. \label{Tsvsw}
\end{equation}

The following table summarizes the previous discussion about the space of solutions of the wedge condition together with the algebra structure that we discuss in subsection \ref{Sec:Walgebra}.

\begin{table}[H]
\centering
\begin{tabular}{c|cccccc|l}
\toprule
manifold & $\T{-1}$ & $\T0$ & $\T1$ & $\T2$ & $\cdots$ & $\T{s}$ & \qquad algebra\\
\midrule
$S_0$   &  3  &  4  &  3  & 0 & $\ldots$ & 0   & $ \Wcal(S_0)=  \mathfrak{iso}_-(3,1)\,\oplus\, \C^3$ \\
$S_1$   &  1  &  2  &  3  & 4 & $\ldots$ & s+2 & $\Wcal(S_1)= L_+(w_{1+\infty})$\\
$S_2$   &  1  &  2  &  3  & 4 & $\ldots$ & s+2 & $\Wcal(S_2)= Lw_{1+\infty}$ \\
\bottomrule
\end{tabular}
\caption{\textit{Number of independent complex solutions to $D^{s+2}\T{s}=0$ depending on the choice of manifold. For $S_1$ and $S_2$, each subspace of degree $s$ has complex dimension $(s+2)\cdot\infty$ due to the loop part.} \label{table1}}
\end{table}

\subsection{The $\W$-bracket \label{sec:Wbracket}}

We equip $\W(S)$ with the following $\W$-bracket:
\be 
[T,T']^\W_s := \sum_{n=0}^{s+1} (n+1)\left( \T{n} D \Tp{s+1-n}-\Tp{n} D \T{s+1-n}\right). \label{Wbracket}
\ee
Notice that the bracket acts on the family $T= (T_s)_{s+1\in \N}$ rather than on a single object $\T{s}$ with definite helicity $-s$.
This notation will turn out to be very convenient and allow us to derive statements valid for any spin-weights. Quite remarkably, this bracket is defined on $\W(S)$ independently of the topology of $S$.

This bracket is usually written in term of its components. 
This means that  $\big[\iota(\T{p}),\iota(\Tp{q})\big]_{s}$  vanishes unless $s=p+q-1$. 
Hence one could also write \eqref{Wbracket} as
\begin{equation}
    \big[\iota(\T{p}), \iota(\Tp{q})\big]_{p+q-1}^\W=(p+1)\T{p} D\Tp{q}-(q+1)\Tp{q} D\T{p}, \label{Wbracketbis}
\end{equation}
which shows that $[\cdot\,,\cdot]^\W$ is a bracket of degree $-1$, namely
\begin{equation}
    [\cdot\,,\cdot]^\W: \W_p\times\W_q\to\V_{p+q-1}.
\end{equation}
In our case, we choose to define the natural extension of the latter to the full graded space $\W(S)$, namely
\begin{equation}
    [\cdot\,,\cdot]^\W: \W(S)\times\W(S)\to\V(S),
\end{equation}
such that we take the sum over all combinations of $p$ and $q$ with $p+q-1=s$ to recover \eqref{Wbracket}:
\begin{equation}
    \boxed{\big[T,T'\big]^\W_s= \sum_{p+q=s+1}\big[\iota(\T{p}),\iota(\Tp{q})\big]_{p+q-1}^\W}.
\end{equation}

An important property of this bracket (which is the reason why we use the superscript $\W$) is that it closes on the wedge. 
In other words, $D^{s+2}[T,T']_s=0$ when both $T$ and $T'$ are in $\W(S)$. 
Hence we get (see the proof in Appendix \ref{AppProofWedge})
\begin{equation}
[\cdot\,,\cdot]^\W:\W(S)\times\W(S)\to\W(S). \label{Wbracketclosure}
\end{equation}

\ni \textbf{Relation between $T_s$ and the celestial modes $w^p_m$:}\\
Defining the modes $w^p_{m,k}=-\iota\left(\frac12 z^{p-1+m}\bz^k \right)$, cf.\,\eqref{Tsvsw}, we find that the $\W$-bracket of the latter reduces to 
\begin{equation}
    \big[ w^p_{m,k}, w^q_{n,\l}\big]^\W_{2(p+q-2)-3}=\big(m(q-1)-n(p-1)\big)w^{p+q-2}_{m+n,k+\l}.
\end{equation}
This is the usual form of the $Lw_{1+\infty}$ Lie bracket.

\subsection{The $\Wcal$-algebra \label{Sec:Walgebra}} 

We are now ready to show the following lemma.

\begin{tcolorbox}[colback=beige, colframe=argile] \label{LemmaWedgeAlg}
\textbf{Lemma [Wedge algebra]}\\
The wedge algebra $\Wcal(S)\equiv\big(\W(S),[\cdot\,,\cdot]^\W\big)$ equipped with the bracket \eqref{Wbracket} forms a Lie algebra. 
\end{tcolorbox}

\ni As we already emphasized in the table \ref{table1}, $\Wcal(S)$ is a \emph{generalization} of $Lw_{1+\infty}$ since it is defined for any 2d surface $S$ of arbitrary topology, while  $Lw_{1+\infty}$ is equal to $\Wcal(S_2)$.

\paragraph{Proof:}  
The linearity and anti-symmetry of \eqref{Wbracket} are obvious. 
Besides, we already discussed the closure of the bracket in \eqref{Wbracketclosure}.
In Appendix \ref{AppJacobiVbracket} we also show that 
\be\label{Jac1}
\big[T,[T',T'']^\W\big]_s^\W \cyc (s+3) D^2\T0\big(\Tp0 \Tpp{s+2} - \Tp{s+2} \Tpp0\big).
\ee
The symbol $\cyc$ means that we add the cyclic permutation of $(T,T',T'')$ on both sides of the equal sign, i.e.
\begin{align}
    & f(T,T',T'')\cyc g(T,T',T'') \\
    \Leftrightarrow \quad & f(T,T',T'')+f(T',T'',T)+f(T'',T,T')=g(T,T',T'')+g(T',T'',T)+g(T'',T,T'), \nn
\end{align}
for some arbitrary functionals $f$ and $g$.
It means in particular that
\begin{equation}
    f(T,T',T'')\cyc f(T',T'',T)\cyc f(T'',T,T').
\end{equation}
We see that the RHS of \eqref{Jac1} vanishes when $D^2T_0=0$ which is implied  by the wedge condition.\footnote{The closure of the algebra requires the wedge condition $D^{s+2}\T{s}=0$ for all $s\geqslant -1$ (and not only for $s=0$).} And therefore this shows that the Jacobi identity is satisfied when $T,T',T''\in \W(S)$. 

\paragraph{Twistorial proof:}
We present for completeness an alternative proof of the \hyperref[LemmaWedgeAlg]{Lemma} which highlights the relationship with twistor space when $S=S_2$.
Indeed, the $\W$-bracket satisfies the Jacobi identity since it originates from the image of $L(\mathrm{Ham}(\mathbb{C}^2))$ when projected from twistor space to $\scri$. 
We refer the reader to \cite{Donnay:2024qwq}, whose notation we follow, for extra details.
Projective twistor space can be written as an $O(1)\oplus O(1)$ bundle over $\mathbb{CP}_1$. 
The anti-holomorphic fibers coordinates of this bundle are represented as coordinate $\bar\mu^{\alpha}$ on $\C^2$.
A hamiltonian generator of area preserving diffeomorphism  (i.e. with unit Jacobian\footnote{Which amounts to require the tracelessness of the symmetry parameters. 
This is trivial in twistor space since using \eqref{twistorbracket}, we know that $\poisson{g,\cdot}=n\,g_{\alpha\alpha_2\ldots\alpha_n}(\bz)\bar\mu^{\dot\alpha_2}\ldots\bar\mu^{\alpha_n}\epsilon^{\alpha\beta}\frac{\pa}{\pa\bar\mu^\beta}$. $g_{\alpha(n)}$ being symmetric, we immediately get that the symmetry parameter is trace-free, i.e. $n\,g_{\alpha\beta\alpha_3\ldots\alpha_n}\epsilon^{\alpha\beta}=0$.}) of degree $s+1$ is given by 
\begin{equation}
    g_s=g_{\alpha(s+1)}(\bz)\, \bar\mu^{\alpha(s+1)}, \label{HamGen}
\end{equation}
where $\alpha(n)$ is a multi-index notation that stands for\footnote{$\bar\mu^{\alpha(n)}=\bar\mu^{\alpha_1}\ldots \bar\mu^{\alpha_n}$.} $\alpha_1\cdots\alpha_n$ and  where $g_{\alpha(s+1)}(\bz)$ is  an anti-holomorphic function on $S_2$, 
\begin{equation}
    g_{\alpha(s+1)}(\bz)=\sum_{k=-\infty}^\infty g_{\alpha(s+1)}^{(k)}\bz^k.
\end{equation}
We consider another hamiltonian generator $g'_{s'}$ of degree $s'+1$,
\begin{equation}
    g'_{s'}=g'_{\alpha(s'+1)}(\bz)\, \bar\mu^{\alpha(s'+1)}.
\end{equation}
The algebra $Lw_{1+\infty}$ is then realized through the Poisson bracket
\begin{equation}
    \{\cdot\,,\cdot\}= \epsilon^{\alpha\beta}\frac{\pa}{\pa\bar\mu^\alpha}\frac{\pa}{\pa\bar\mu^\beta}, \label{twistorbracket}
\end{equation}
so that
\begin{equation}
    \{g_s,g'_{s'}\}=g''_{s+s'-1}=g''_{\alpha(s+s')}(\bz)\,\bar\mu^{\alpha(s+s')},
\end{equation}
where
\begin{equation}
    g''_{\alpha(s+s')}(\bz)=(s+1)(s'+1)\epsilon^{\alpha\beta} g_{\alpha\alpha_1\ldots\alpha_s}(\bz)g'_{\beta\alpha_{s+1}\ldots\alpha_{s+s'}}(\bz).
\end{equation}
Still following the notation of \cite{Donnay:2024qwq}, we project the algebra down to $\scri$.
The hamiltonian generator $g_s$ is mapped to the function $\tT_s(\lambda^\alpha,\bz)$ and then to the function $\T{s}(z,\bz)$ given by  
\begin{equation}
    g_s \quad\longrightarrow\quad \tT_s(\lambda^\alpha,\bz)=g_{\alpha(s+1)}(\bz)\lambda^{\alpha(s+1)}=(-\lambda^1)^{s+1}\T{s}(z,\bz), \label{HamGenScri}
\end{equation}
where $z\equiv-\lambda^0/\lambda^1=\lambda_1/\lambda_0$. The function $T_s$ is such that 
\begin{equation}
    \T{s}(z,\bz):=g_{\alpha(s+1)}(\bz)\frac{\lambda^{\alpha(s+1)}}{(-\lambda^1)^{s+1}}=\sum_{k=0}^{s+1}\binom{s+1}{k}(-1)^{s+1-k}z^k\, g_{0(k)1(s+1-k)}(\bz). \label{Tstwistor}
\end{equation}
We define similarly $\tT'_{s'}(\lambda^\alpha,\bz)$ and $\Tp{s}$ for $g'_{s'}$.
Besides, the Poisson bracket \eqref{twistorbracket} is sent to
\begin{equation}
    \{\cdot\,,\cdot\}^{\scri}= \epsilon^{\alpha\beta}\frac{\pa}{\pa\lambda^\alpha}\frac{\pa}{\pa\lambda^\beta}. \label{twistorbracketscri}
\end{equation}
The easiest way to see the correspondence with the $\W$-bracket is to evaluate the bracket in terms of $\pa_z$ acting on $T_s$:
\begin{align}
    \poisson{\tT_s,\tT'_{s'}}^{\scri} &=\epsilon^{\alpha\beta}\left(
    (-\lambda^1)^{s+1}\frac{\pa z}{\pa\lambda^\alpha}\pa_z\T{s}+\T{s}\frac{\pa(-\lambda^1)^{s+1}}{\pa\lambda^\alpha}\right)\left((-\lambda^1)^{s'+1}\frac{\pa z}{\pa\lambda^\beta}\pa_z\Tp{s'}+\Tp{s'}\frac{\pa(-\lambda^1)^{s'+1}}{\pa\lambda^\beta}\right) \nn\\
    &=(-\lambda^1)^{s+s'}\epsilon^{\alpha\beta} \big(\pa_z\T{s}(\delta^0_\alpha+z\delta^1_\alpha)-(s+1)\T{s}\delta^1_\alpha\big)\big(\pa_z\Tp{s'}(\delta^0_\beta+z\delta^1_\beta)-(s'+1)\Tp{s'}\delta^1_\beta\big) \nn\\
    &=(-\lambda^1)^{s+s'}\big((s+1)\T{s}\pa_z \Tp{s'}-(s'+1)\Tp{s'}\pa_z\T{s}\big).
\end{align}
From there we identify\footnote{On $\mathbb{C}^*$ we have $D =\pa_z$.}
\begin{equation}
    \Tpp{s+s'-1}\equiv (s+1)\T{s}\pa_z \Tp{s'}-(s'+1)\Tp{s'}\pa_z\T{s} =[\iota(\T{s}),\iota(\Tp{s'})]_{s+s'-1}^\W.
\end{equation}
The $\W$-bracket thus descends from the canonical Poisson bracket on twistor space, which proves the isomorphism between the projection of $L(\mathrm{Ham}(\C^2))$ on $\scri$ and the $\Wcal$-algebra.
Notice in particular that since $D g_{\alpha(s)}=0$ we have that  $\T{s}$ given by \eqref{Tstwistor} satisfies  $D^{s+2}\T{s}=0$.
The fact that the Jacobi identity holds for $[\cdot\,,\cdot]^\W$ is thus merely a consequence of $\{\cdot\,,\cdot\}$ satisfying Jacobi. 

We bring the reader's attention to the shift in degree between the twistor perspective compare to the gradation in $\V(S)$.
Indeed, the element $\T{s}\in\V_s$ originates from a polynomial of degree $s+1$ in $\bar\mu^\alpha$ or $\lambda^\alpha$ in the twistor perspective (cf. \eqref{HamGen} and \eqref{HamGenScri}).
In particular, the spin 0 super-translations are polynomials of degree 1 in $\bar\mu^\alpha$, so that for $s=0=s'$, we get $\poisson{g_\alpha\bar\mu^\alpha,g'_\beta\bar\mu^\beta}=\epsilon^{\alpha\beta}g_\alpha g'_\beta\neq 0$.
In other words, the super-translations do not commute in twistor space.
From the $\Wcal$-algebra viewpoint, this amounts to the presence of central elements of degree $-1$, as we now discuss.

\paragraph{Central elements:}
Notice that the space  $\W_{-1}(S)$ is central for the wedge bracket. 
Given $c\in \W_{-1}$, we denote by $\hat{c}:=\iota(c) \equiv(c,0,\ldots)$ the series for which $\pi_{-1}(\hat c)=c$ and $\pi_{s}(\hat c)=0$ for $s\geqslant 0$.
It is easy to see that $\hat c$ commutes with any element of $\Wcal$,
\begin{equation}
    [\hat c,T]_s^\W =0,\qquad\forall \,T\in\W(S),\quad s\geqslant -1. \label{central-1}
\end{equation}
The fact that $\W_{-1}(S)$ is central and thus forms an ideal, implies that we can define the quotient algebra 
$\bWcal(S)\equiv\big(\Ws,[\cdot\,,\cdot]^{\overline\W}\big)$ where 
\be
\Ws = \W(S)/\W_{-1}(S) \simeq  \bigoplus_{s=0}^{\infty } \W_s(S). 
\ee 

To define the quotient bracket $[\cdot\,,\cdot]^{\overline\W}$ one recalls that, since the bracket is of degree $-1$, the only way to obtain a degree $-1$ element is through the commutation of two supertranslations, namely
\be 
[T,T' ]^{\W}_{-1}=\T0 D\Tp0-\Tp0 D\T0~ \in \W_{-1}(S).
\ee 
This means that the supertranslations do not commute in $\Wcal(S)$ while they do in $\overline\Wcal(S)$. Overall we have that 
\be 
[T,T' ]^{\Wo}_{-1}=0, 
\qquad
[T,T' ]^{\Wo}_{s}= [T,T' ]^{\W}_s, \quad \mathrm{for}\quad s\geq 0.
\ee 
Therefore all the brackets
in $\bWcal(S)$ agree with the ones in $\Wcal(S)$ except for  the bracket of two supertranslations.

\paragraph{Wedge and Poincaré algebra:}
It is interesting to understand how the central extension works in the case of $S=S_0$ where $\bWcal(S)$ is the self dual truncation of the Poincaré algebra and $\Wcal(S)$ is a three dimensional central extension of it.
As we discussed, we have that $\W(S_0) = \W_{-1}(S_0)\oplus \W_{0}(S_0)\oplus\W_{1}(S_0)$ is ten dimensional.
The basis elements are parametrized by
\be 
C_i := D n_i \in \W_{-1}(S_0), \qquad 
P_{a} := q_a \in \W_0(S_0), 
\qquad
\bJ_i  := \bD n_i \in \W_{1}(S_0),
\ee 
where $i=1,2,3$ and $a=0,1,2,3$, while $q(z,\bz)= (1, n(z,\bz))$ was defined in \eqref{T0T}.
From the definition of the wedge algebra bracket, we can compute the algebra between these generators.
We find that $C_i$ commutes with $(C_j,P_a,\bJ_j)$, while the non trivial commutators are 
\bs \label{PoincB}
\be 
[\bJ_i,\bJ_j] &= -2i\epsilon_{ij}{}^k \bJ_k,\\
[\bJ_i,P_b] &= \bar\eta_{abi}P^a,\\
[P_a ,P_b ] &= \eta_{a b}{}^i C_i,
\ee
\es
where $\epsilon_{ijk}$ is the Levi-Civita tensor, while 
$\eta_{ab}{}^i$ and $\bar\eta_{ab}{}^i$ are the Lorentzian versions of the t'Hooft symbols \cite{t1976computation}. 
They are antisymmetric in $(a,b)$ and their components are given by 
\be 
\eta_{ij}{}^k =\bar\eta_{ij}{}^k= i \epsilon_{ij}{}^k, \qquad 
\eta_{0i}{}^k =-\bar\eta_{0i}{}^k= \delta_i{}^k. \label{tHooftSymbol}
\ee
$\eta_{ab}{}^i$ and $\bar\eta_{ab}{}^i$ are respectively projectors onto the self-dual and anti-self-dual basis of 2-forms.\footnote{One has $(\star\,\eta)_{ab}{}^k=\frac12\epsilon_{ab}{}^{cd} \eta_{cd}{}^k= i \eta_{a b}{}^k$ and $(\star\,\bar\eta)_{ab}{}^k=\frac12\epsilon_{ab}{}^{cd} \bar\eta_{cd}{}^k= -i\bar\eta_{a b}{}^k$, where $\star$ is the Hodge star associated to the Minkowski metric.}
In other words, the three generators $\bJ^k$ are nothing else than the anti-self-dual part of the usual Lorentz generators $\mathcal{J}^{ab}$, i.e. $\bJ^k=\frac{i}{2}\bar\eta_{ab}{}^k\mathcal{J}^{ab}$.
The quotient algebra $\overline{\mathcal{W}}(S_0)$ is obtained by imposing $C_i=0$. In this case we get the anti-self-dual Poincaré algebra
\be 
\overline{\mathcal{W}}(S_0)
= \mathfrak{sl}_{-}(2,\C)\sds \C^4\equiv \mathfrak{iso}_-(4),
\ee 
where $ \mathfrak{sl}(2,\C)_{-}$ denotes the anti-self-dual projection of the complexified Lorentz algebra.
Although the Poincaré algebra $\big(\mathfrak{sl}_{-}(2,\C)\oplus\overline{\mathfrak{sl}}_{+}(2,\C)\big)\sds\C^4$ does not admit any central extension \cite{Nakayama:2023xzu}, it is remarkable that its (anti-)self-dual part does.
The proof of the brackets \eqref{PoincB} is given in appendix \ref{app:Poinc} and relies on the validity of the following identities:
\be \label{nids}
D^2 n_i &=0, & 2 D n_i \bD n_j + n_i n_j &= \delta_{ij} + i \epsilon_{ij}{}^k n_k,\cr
D\bD n_i &= - n_i, &
n_i D n_j - n_j D n_i &= i \epsilon_{ij}{}^k D n_k.
\ee
It is interesting to note that the centrality of $C_i$ is consistent with the Jacobi identity.\footnote{Requiring $[\bJ_i,[P_a,P_b]]=0$ demands that  
$\eta_{b}{}^{cj}\bar\eta_{cai} - \eta_{a}{}^{cj}\bar\eta_{cbi} =0$, which is satisfied due to the orthogonality of self-dual and anti-self-dual projections.}

The central extension of the chiral Poincaré algebra can also be written in spinor variables. In this case the different generators are $(\bJ_{\dot\alpha\dot\beta}, P_{\dot\alpha\alpha}, C_{\alpha \beta})$ which renders manifest the respective chirality of each generator.\footnote{$\bJ_{\dot\alpha\dot\beta}$ is symmetric in $(\dot\alpha,\dot\beta)$ and similarly $C_{\alpha \beta}$ is symmetric.} The translation commutator is given by 
\be 
[P_{\dot\alpha \alpha}, P_{\dot\beta \beta}]= \epsilon_{\dot\alpha\dot\beta} C_{\alpha \beta}.
\ee 
This expression makes it clear that while the anti-self-dual rotation $\bJ_{\dot\alpha\dot\beta}$ acts on $P_{\dot\alpha\alpha}$ on the left, it does not act on $C_{\alpha\beta}$, respecting its chirality. 
Note that the dual chiral algebra is given by generators 
$(J_{\alpha \beta}, \bP_{\alpha \dot\alpha}, \bC_{\dot\alpha\dot\beta})$, where $\bP_{\alpha \dot\alpha}$ denotes the opposite chirality translation generators that commute with $\bJ$ and transform as a vector under the action of $J$.

To recover the usual action of Poincaré one needs to impose the reality condition 
$\bP_{a}=P_{a}\equiv\mathcal{P}_a$. 
This condition means that $[J_i, \mathcal{P}_b]=\eta_{abi}\mathcal{P}^a$ and 
$[\bJ_i, \mathcal{P}_b]=\bar\eta_{abi}\mathcal{P}^a$. The Jacobi identity then implies that $[J_i,[\mathcal{P}_a,\mathcal{P}_b]]\neq 0$. 
Hence  the only way to respect the centrality of $C$'s is to impose $C_{i}=\bC_{i}=0$.\footnote{Another way to be consistent with Jacobi is to impose $C_i =\Lambda J_i$ and $\bC_i=\Lambda \bJ_i$. The $C$'s are no longer central and we recover the de Sitter or anti-de Sitter algebra depending on the sign of $\Lambda$.} 
With these 10 constraints we have that 
$\mathcal{P}_a, J_i, \bJ_i$ represent $4+3+3=10$ generators which form the Poincaré algebra.
We learned \textit{a posteriori} that this central extension of chiral Poincaré was already discussed in \cite{Krasnov:2021cva}.

\section{Deforming the wedge algebra \label{sec:DeformingWedge}}

In this section we show that the presence of a shear element $\sigma \in \Ccel{(1,2)}$ allows for the deformation of the wedge algebra.

\subsection{Can we extend the $\W$-bracket beyond the wedge?}

We know that the $\gbms$ algebra closes without any wedge restriction on $\T0$ and $\T1$.
Besides, it was proven in \cite{Freidel:2023gue} that the wedge condition on $\T2$ is not necessary for the algebra of higher-spin charges to close at the quadratic level. In the same work, it was also proven that, in Yang-Mills, the wedge restriction can be lifted at quadratic order for all spin $s$ charges.
These results strongly suggest that the symmetry can be extended beyond the wedge and we now investigate how we can waive the restriction $D^{s+2}\T{s}=0$ at the algebraic level. 

The first question that arises is whether we can simply consider the bracket $[\cdot\,,\cdot]^\W$ 
to be valid beyond the wedge, i.e. to be viewed as a map
$[\cdot\,,\cdot]^\V:\V(S)\times\V(S)\to\V(S)$. This means that
\begin{equation}
    [T,T']^\V_s= \sum_{n=0}^{s+1} (n+1)\big(\T{n} D \Tp{s+1-n}-\Tp{n} D \T{s+1-n}\big),  \label{Vbracket}
\end{equation}
for $T,T'\in\V(S)$. 
This $\V$-bracket is simply the naive extension of the $\W$-bracket. 
As we are about to see, it is ill-defined since fails to satisfy the Jacobi identity, once we go beyond the wedge.
This is surprising at first sight since the $\V$-bracket is very closely related to a well-known bracket: 
the Schouten-Nijenhuis (SN) bracket for \textit{symmetric} contravariant tensor fields.
Be mindful that the literature often deals with the SN bracket for \textit{antisymmetric} multivectors. 
For the symmetric version, check the original reference \cite{Schouten} and \cite{SNbracket} for a modern perspective.

\begin{tcolorbox}[colback=beige, colframe=argile, breakable] \label{SNreminder}
\textbf{Reminder [Schouten-Nijenhuis bracket]}\\
Consider $\X^\bullet(M)= \bigodot^\bullet\Gamma(TM)$, the algebra of symmetric contravariant tensor fields of any rank over a $d$ dimensional manifold $M$ with coordinates $\{x^i\}_{i=1}^d$. For $S_p\in\X^p(M)$ and $\tilde S_q\in\X^q(M)$, the SN bracket is a graded commutator of degree 1, i.e.
\begin{center}
\begin{math}
    [\cdot\,,\cdot]^\sn:\X^p(M)\times \X^q(M)\to\X^{p+q-1}(M),\quad \forall\,p,q\in\N,
\end{math}
\end{center}
with components given by
\begin{equation}
    \big[S_p,\tilde S_q\big]^{\sn, i_1\ldots i_{p+q-1}}=pS_p^{j(i_1\ldots i_{p-1}}\frac{\pa}{\pa x^j}\tilde S_q^{i_{p} \ldots i_{p+q-1})}-
    q\tilde S_q^{j(i_1\ldots i_{q-1}}\frac{\pa}{\pa x^j} S_p^{i_q\ldots i_{q+p-1})}. \label{SNbracketDef}
\end{equation}
This is a well-defined tensorial expression, so that the partial derivatives can be replaced by any  torsion-free covariant derivative.
In our case where the manifold is the 2-sphere/complex plane (such that $i_1,i_2\ldots\to A_1,A_2\ldots$ and $\{x^A\}\equiv(z,\bz)$) and we restrict to symmetric \textit{holomorphic traceless} tensors, the projection of \eqref{SNbracketDef} along $\bm_{A_1}\ldots\bm_{A_{p+q-1}}$ reads\footnote{We use that the sphere metric $\gamma_{AB}=m_A\bm_B+\bm_A m_B$.}${}^,$\footnote{Notice the natural abuse of notation where on the LHS, $S_p$ refers to the full tensor as in \eqref{SNbracketDef}, while on the RHS, it stands for the projected quantity along $\bm$'s.}
\begin{equation}
    \big[S_p,\tilde S_q\big]_{p+q-1}^\sn=pS_p D\tilde S_q-q\tilde S_q DS_p.
\end{equation}
This means that for generic $S,\tilde S\in\X^\bullet(M)$ we have 
\begin{equation}
    \boxed{\big[S,\tilde S\big]^\sn_s= \sum_{p+q=s+1}\big[S_p,\tilde S_q\big]_{p+q-1}^\sn=\sum_{n=0}^{s+1} n\big(S_nD\tilde S_{s+1-n} -\tilde S_nDS_{s+1-n}\big).} \label{SNbracketDefbis}
\end{equation}
\end{tcolorbox}
We thus see that what we are calling $\V$-bracket is actually a shifted SN bracket where the prefactor between \eqref{SNbracketDefbis} and \eqref{Vbracket} is shifted from $n$ to $n+1$.
Now comes the intriguing part, with far reaching consequences: Although the original SN bracket satisfies the Jacobi identity (see App.\,\ref{JacobiSN} for a reminder), this is not the case for \eqref{Vbracket}.

In the appendix \ref{AppJacobiVbracket}, we  demonstrate that
\begin{equation}
    \boxed{\left[T,[T',T'']^\V\right]^\V_s \cyc D^2\T0 \paren{T',T''}_s}, \label{Jacobianomaly}
\end{equation}
where we introduced the new quantity\footnote{NC refers to $\paren{\cdot\,,\cdot}$ as Dali's bracket, as a reference to the paintings with melting watches.}
\begin{equation}
\boxed{\paren{T,T'}_s:=(s+3) \left( \T0\Tp{s+2}-\Tp0\T{s+2}\right)} \quad\in ~{\Ccel{(-2,-s-2)}}. \label{Cbracket}
\end{equation}
Notice that the Jacobi anomaly is proportional to $D^2\T0$.
In other words, it disappears if we restrict the algebra to the wedge, as it should.

\paragraph{Relation to $\gbms$:}
If we consider solely transformations of helicity $s=0$ and $s=1$, for which $\T{-1}\equiv 0$ and $\T{s+2}\equiv 0$ for $s\geqslant 0$, then \eqref{Cbracket} vanishes. 
Formally, this amounts to $T,T'\in\V_0\oplus\V_1$.
The bracket \eqref{Vbracket} restricted to these symmetry parameters is explicitly given by
\bs
\label{bracketdeg01}
\be
[T,T']^\V_0 &= (\T0 D\Tp1 - \Tp0 D\T1)  + 2(\T1 D\Tp0 - \Tp1 D\T0), \label{bracketdeg0}\\
[T,T']^\V_1 &= (\T0 D\Tp2 - \Tp0 D\T2)  + 2(\T1 D\Tp1 - \Tp1 D\T1). \label{bracketdeg1}
\ee
\es
The quotient algebra obtained by imposing the vanishing of the central elements, i.e. $\T{-1}=0$ is the $\gbms$ algebra \cite{Campiglia:2014yka, Compere:2018ylh, Campiglia:2020qvc, Freidel:2021fxf}.
To recognize the latter, we denote $\T0\equiv T$, the super-translations parameter and $\T1\equiv Y/2$, the super-Lorentz-rotations parameter. 
We get that the super-translations commute,
\begin{subequations}
\begin{equation}
   [T,T']^{\gbms}=0. \label{gbmstranslation}
\end{equation}
Picking respectively $\T0=0,\T1=Y/2,\Tp0=T', \Tp1=0$ in \eqref{bracketdeg0} and $\T0=0,\T1=Y/2,\Tp0=0, \Tp1=Y'/2$ in \eqref{bracketdeg1}, we get the commutation relations with super-rotations:
\be 
    [Y, T']^{\gbms}= Y DT'-\frac12 T' DY,\qquad [Y,Y']^{\gbms} = YDY'-Y'D Y. \label{gbmsrotation}
\ee 
\end{subequations}
We employ the notation $[\cdot\,,\cdot]^{\gbms}$ which represents  $[\iota(\cdot)\,,\iota(\cdot)]^{\overline{\V}}$  restricted to spin $0$ and $1$, where $\Vs$ represents the quotient $\V(S)/\V_{-1}$.
Importantly, $\gbms$ forms an algebra without any wedge restriction $D^2 T=0=D^3 Y$ on its parameters.\\

There is even a third case where \eqref{Cbracket} cancels out. 
Indeed, even if the anomalous quantity depends on arbitrary high spins $s$, it also always includes $\T0$.
By restricting the algebra to $\Vss$ (where precisely $\T0\equiv 0$, cf. \eqref{VstarstarDef}), we thus restore the Jacobi identity.
On top of that, notice that the $\V$-bracket closes on $\Vss$,\footnote{It is clear since for any $T,T'\in\Vss$, we have that $[T,T']_0^\V = 0$. Hence $[T,T']\in \Vss$ .}
\begin{equation}
[\cdot\,,\cdot]^\V:\Vss\times\Vss\to\Vss.
\end{equation}
We thus proved that 
\begin{tcolorbox}[colback=beige, colframe=argile]
\textbf{Lemma [$\bbVcal$ algebra]}\\
The space $\bbVcal(S)\equiv \big(\Vss,[\cdot\,,\cdot]^\V\big)$ forms a Lie algebra.
\end{tcolorbox}
This Lie algebra is a higher spin generalization of $\mathrm{Diff}(S)$ which is contained in $\bbVcal(S)$ as its spin $1$ component. 
This proves that there exist three algebras, $\Wcal(S),\,\gbms(S)$ and $\bbVcal(S)$, for which the $\V$-bracket is a genuine Lie bracket.\footnote{This includes any sub-algebras of these, such as the $\ebms(S)$ sub-algebra of $\gbms(S)$ which consists of holomorphic vector fields on $S$, i.e. vector fields satisfying $\bD Y=0$. }
The next step consists in building a new bracket, denoted $[\cdot\,,\cdot]^\sigma$, that reduces to the $\V$-bracket for $\Wcal$, $\gbms$ and $\bbVcal$.

\subsection{Covariant wedge algebra \label{sec:NewFieldDepBracket}}

In order to resolve the Jacobi identity issue \eqref{Jacobianomaly}, we introduce a new field dependent bracket 
\begin{empheq}[box=\fbox]{align}
    [T,T']^\sigma_s &=[T,T']^\V_s -\sigma\paren{T,T'}_s,\qquad s\geqslant -1\nn\\
    &=\sum_{n=0}^{s+1}(n+1)\big(\T{n}D\Tp{s+1-n}-\Tp{n}D\T{s+1-n}\big)-(s+3)\sigma\big(\T0\Tp{s+2}-\Tp0\T{s+2}\big), \label{CFbracket}
\end{empheq}
where $\sigma(z,\bz)$ is a spin-weighted function on $S$.
At that stage, it can be viewed as a parameter of the algebra. 
The physical motivation for this deformed bracket is given in \cite{Cresto:2024mne}.
From a purely algebraic point of view, notice that the $\sigma$-bracket is built from the $\V$-bracket by subtracting the part of the Jacobi identity anomaly which looks like a bracket, namely what we called the Dali-bracket $\paren{\cdot\,,\cdot}$.
The latter is multiplied by the deformation parameter $\sigma$, whose properties we constraint in order to construct a Lie bracket.

The difference between $[\cdot\,,\cdot]^\sigma$ and $[\cdot\,,\cdot]^\V$ involves $\paren{T,T'}_s$, which vanishes for $\gbms$
and $\Vss$. Therefore, $[\cdot\,,\cdot]^\sigma$ reduces to the $\V$-bracket for these cases.
Note that $\paren{T,T'}_s$ has celestial weight $-2$ and helicity $-(s+2)$; therefore the additional field dependent term $\sigma\paren{T,T'}_s$ belongs to $\V_s$ provided we choose $\sigma$ of celestial/spin weights $(1,2)$.\footnote{This implies that $\pi_s (\sigma \paren{T,T'})= \sigma \paren{T,T'}_s $.}

Let us investigate the Jacobi identity for that newly proposed bracket. 
The double commutator expands as
\begin{align}
     \big[T,[T',T'']^\sigma\big]_s^\sigma
     &= \big[T,[T',T'']^\V\big]_s^\V
     - \sigma \paren{T,[T',T'']^\V}_s
     - \big[T, \sigma\paren{T',T''}\big]_s^\V
     + \sigma \paren{T, \sigma\paren{T',T''}}_s.
\end{align}
We then evaluate the cyclic permutation of each term entering this expression. 
We report the details of computation in appendix \ref{AppJacobiCF}.
Besides \eqref{Jacobianomaly}, we have 
\begin{equation}
    - \sigma \paren{T,[T',T'']^\V}_s -\big[T,\sigma\paren{T',T''}\big]_s^\V \cyc -\big(2D\sigma\T1+3\sigma D\T1\big)\paren{T',T''}_s \label{Jacobi1}
\end{equation}
and
\begin{equation}
    \sigma \paren{T, \sigma \paren{T',T''}}_s \cyc 3 \sigma^2\T2\paren{T',T''}_s. \label{Jacobi2}
\end{equation}
Therefore we conclude that
\begin{align}
     \boxed{\big[T,[T',T'']^\sigma\big]_s^\sigma\cyc -\hdT \sigma\paren{T',T''}_s}, \qquad s\geqslant -1,\label{JacCF}
\end{align}
where
\begin{equation}
    \boxed{\hdT \sigma :=-D^2\T0+2D\sigma\T1+3\sigma D\T1-3\sigma^2\T2} \quad \in\, \Ccel{(1,2)}. \label{dTCv1}
\end{equation}

In order for the Jacobi identity to hold, the simplest choice is to set $\hdT \sigma=0$, a condition under which $[\cdot\,,\cdot]^\sigma$ is a well-behaved Lie algebra bracket.
In that case, $\sigma$ is a  field whose field variation vanishes.\footnote{It is a functional constant in the sense that $\hdT \sigma=0$.}
It plays the role of a background structure that represents the symmetry algebra atop a coherent state of gravitons. 
As we will see in more details, $\sigma$ also represents a deformation parameter mapping the wedge algebra onto the generalized wedge algebra $\Wcal_{\sigma}(S)$.
Note that the condition $\hdT\sigma=0$ only gives a relation between $(\T0,\T1,\T2)$. 
However if we demand that such elements form an algebra, i.e. that $\hat\delta_{[T,T']^\sigma}\sigma=0$ when $\hdT\sigma=0=\hdTp\sigma$, then this leads to a generalization of the wedge condition on the parameters $T_s$.
To linear order in $\sigma$, this generalized wedge condition reads 
\be \label{wedge approximate}
0= D^{s+2} T_s -\sum_{k=0}^{s+1} (k+1) D^k\big(\sigma D^{s+1-k}T_{s+1}\big) +\cdots,\qquad s\geqslant -1,
\ee 
where the dots stand for terms of higher powers in $\sigma$ when $s>-1$.\\

We now demonstrate the various statements of the last paragraph. 
For this, it is essential to introduce a preferred element $\Ham \in \V_{-1} \oplus \V_0$.\footnote{Any element of $\V_{-1}$ is central for the $\sigma$-bracket, like $\W_{-1}$ was for the $\W$-bracket.}
In other words $\Ham :=(\Ham_{-1},\Ham_0,0,0\ldots)$ is such that $\pi_s(\Ham)=0,\,s>0$. 
Note that $\Ham_0\in \V_{0}$ is, by definition, of conformal dimension $-1$.\footnote{Like the inverse of the  square root of a density on a 2d manifold.}
It therefore transforms non trivially under sphere diffeomorphisms.
We assume that $\Ham_0\neq 0$.
Quite remarkably, this means by Moser's theorem \cite{moser1965volume} that there exists a diffeomorphism frame such that $\Ham_0$ is constant, i.e. $D\Ham_0=0=\bD\Ham_0$.\footnote{The condition $D\Ham_0=0$ is not preserved by diffeomorphisms: 
under an infinitesimal diffeomorphism generated by the element $\iota(Y)$ with $Y\in \V_1$, we have that  $[\Ham, \iota(Y)]^\sigma_0= \Ham_0 DY \neq 0$.
This is analogous to the statement that the energy is not boost invariant.} 
We choose the scale of $\Ham_0$ to be such that $\Ham_0=1$ in this frame. 
We are then left with the choice of $\Ham_{-1}$. 
This freedom can be parameterized by another spin 0 element $G \in \V_0$ through $\Ham_{-1} =\Ham_0 DG \in \V_{-1}$---we can check that the dimensions match as $DG\in \Ccel{(0,1)}$. 
In the following we fix $G$ to be the \emph{Goldstone} mode associated with $\sigma$. 
In other words, $G$ is such that
\be 
  D^2 G =\sigma. \label{Goldstone}
\ee 
To summarize we chose a frame of reference where $\Ham\equiv\hHam$ is given by 
\be 
  \Ham_0=1, \qquad \Ham_{-1}= DG, \qquad \Ham_s=0, \quad \mathrm{for}~ s\geq 1.
\ee

Plugging in \eqref{CFbracket}, we get that
\begin{equation}
    \boxed{[\hHam,T]^\sigma_s= \Ham_0\big(D\T{s+1} -(s+3)\sigma \T{s+2}\big)}~\in\,\V_s,\qquad s\geqslant -1. \label{HamTbracket}
\end{equation} 
We kept $\Ham_0$ explicit to avoid confusion with the weights counting (since $[\hHam,T]_s^\sigma$ and $DT_{s+1}$ have different weights). 
In the frame we work in $\Ham_0=1$, so we could leave it implicit, but it is useful to keep it to remember that $\Ham_0$ is of conformal weight $-1$. 
This reflects the general fact that any equality between expressions with the same spin can be written in an arbitrary frame using powers of $\Ham_0$ to ensure that the conformal weight matches.  

The quantity in parenthesis in \eqref{HamTbracket} is of fundamental importance in the following.
We thus define
\begin{equation}
    \boxed{(\Dcal T)_s:=D\T{s+1} -(s+3)\sigma \T{s+2}}~\in\,\Ccel{(0,-s)}. \label{CovDer}
\end{equation}
Depending on the context, either \eqref{HamTbracket} or \eqref{CovDer} will turn out to be the most convenient way to write or manipulate expressions.
Both equations eventually carry the same information and can be used interchangeably recalling that\footnote{Notice that $\ad\hHam=\ad\Ham_{\scriptscriptstyle{G'}}$ for two Goldstone fields $G$ and $G'$.}
\begin{equation}
    \boxed{\Ham_0\Dcal T=\ad\hHam(T)=[\hHam,T]^\sigma}.
\end{equation}
We introduced the natural notation 
$\ad: \V(S) \to \mathrm{End}(\V(S))$
for the adjoint map. By definition this map is such that 
$ \ad T( T')=[T,T']^\sigma.
$ It sends an element $T\in \V(S)$ to a linear map in $\V(S)$. 
We can restrict this map to $\W_\sigma$ and anticipating the upcoming result that $[\cdot\,,\cdot]^\sigma$ is a Lie algebra bracket, we find that 
$\ad: \Wcal_\sigma(S) \to \mathrm{Der}(\Wcal_\sigma(S))$, i.e.
$\ad T$, with $T \in \Wcal_\sigma(S)$ maps $\Wcal_\sigma$ onto itself and is a derivation of the $\sigma$-bracket. 

We also point out the difference of notation when we write \eqref{CovDer}.
$D\T{s+1}$ is the concise version of $(DT)_s\equiv\pi_s(DT)$.\footnote{Here the celestial weight is irrelevant, so we just assume that $\pi_s$ is defined the same way as the helicity evaluation at degree $s$, on fields with arbitrary conformal dimension.}
While this convention is not confusing when using $D$, it is not possible for $\Dcal$. It is important to keep the parenthesis when dealing with $(\Dcal T)_s$, because of the extra shift in degree for the part proportional to $\sigma$.\footnote{To say it in another way: the action of $\Dcal$ (and powers thereof) is well-defined only on a graded vector $T\in\V(S)$, and not on a single grade $s$ element $\T{s}$.}

The introduction of $\hHam$ and $\Dcal$ allows us to give a new flavor to the quantity $\hdT\sigma$.
To see this, first notice that \eqref{CovDer} makes sense for $s=-2$, even if the $\sigma$-bracket \eqref{HamTbracket} is only defined for $s\geqslant -1$.
In the following, we thus denote
\begin{equation}
    [\hHam,T]^\sigma_{-2}:=\Ham_0(\Dcal T)_{-2}\qquad\textrm{with}\qquad (\Dcal T)_{-2}=D\T{-1}-\sigma\T0. 
    \label{DefAdjoint-2}
\end{equation}
Using this notation we can conveniently write the variation of $\sigma$ as a double commutator:\footnote{Strictly speaking we have $\big[\hHam,[\hHam,T]^\sigma\big]^\sigma_{-2}=-\Ham_0^2 \hdT \sigma$.}
\begin{align}
    \big(\ad^2\hHam(T)\big)_{-2}\equiv \big[\hHam,[\hHam,T]^\sigma\big]^\sigma_{-2}
    &= D[\hHam,T]^\sigma_{-1}- \sigma [\hHam, T]_0^\sigma \cr
    &= D\big(D\T0 - 2 \sigma \T1\big)- \sigma \big(D\T1 -3\sigma \T2\big) \label{doubleB}\\
    &= - \hdT \sigma. \nn
\end{align}
Moreover, since we know the variation $\hdT\sigma$, we can compute $\big[\hdT,\hdTp\big]\sigma$ and $\hat\delta_{[T,T']^\sigma}\sigma$.
The sum of these two reduces to\footnote{The fact that the RHS is non-zero, in other words that $\dT$ is not an algebra action, is connected to the fact that  $[\cdot\,,\cdot]^\sigma$ is not a Lie bracket when $\dT\sigma\neq 0$.}
\begin{align}
    &\big[\hdT,\hdTp\big]\sigma +\hat\delta_{[T,T']^\sigma}\sigma =\T0\,\hat\delta_{[\hHam,T']^\sigma}\sigma -\Tp0\,\hat\delta_{[\hHam,T]^\sigma}\sigma, \label{algebraRepinterm}
\end{align}
where 
\begin{equation}
    -\hat\delta_{[\hHam,T]^\sigma} \sigma =\big(\ad^3\hHam(T)\big)_{-2}:= \left[\hHam,\big[\hHam,[\hHam,T]^\sigma\big]^\sigma\right]^\sigma_{-2}. \label{deltaHamTsigma}
\end{equation}
We report the brute force demonstration of \eqref{algebraRepinterm} in appendix \ref{AppAlgebraAction}.

From  these computations, we deduce that:
$\hdT\sigma$, and thus $(\ad^2\hHam(T))_{-2}$, has to vanish in order for the  Jacobi identity to hold for the $\sigma$-bracket. 
Now given two elements $T,T'\in \V(S)$ such that 
$\hat\delta_T\sigma=\hat\delta_{T'}\sigma=0$ we need to have that $\hat\delta_{[T,T']^\sigma}\sigma=0$ in order for the set of such elements  to form a Lie algebra.

From \eqref{algebraRepinterm} we conclude that this happens if  $ \hat\delta_{[\hHam, T]^\sigma}\sigma$ and $ \hat\delta_{[\hHam, T']^\sigma}\sigma$ vanish too.
To summarize, the property necessary for the Jacobi identity is preserved by the bracket  if $T$ is such that $\hdT\sigma=(\ad^2\hHam (T))_{-2}=0$ and $\hat\delta_{[\hHam, T]^\sigma}\sigma=(\ad^3\hHam(T))_{-2}=0$. Applying the latter to $T=\ad \hHam(T')$ means that we have to demand $(\ad^4\hHam(T))_{-2}=0$ and so on.
This leads us to define the parameter space $\W_\sigma(S)$ as follows.

\begin{tcolorbox}[colback=beige, colframe=argile]
\textbf{Definition [Covariant wedge space]} \label{Wdef}\\
The covariant wedge parameter space is denoted $\W_\sigma(S)$ and given by 
\begin{equation}
\W_\sigma(S):= \Big\{ T \in \V(S)\,~\big|~ \,D\big(\ad^n\hHam(T)\big)_{-1}=\sigma \big(\ad^n\hHam(T)\big)_{0}, \,\forall\,n\geqslant 0 \Big\},
\label{Wdefeq}
\end{equation}
where $\ad\hHam (T):= [\hHam,T]^\sigma$ is the adjoint action of the \textit{Hamiltonian} $\hHam$.
\end{tcolorbox}

\ni This definition is such that if $T\in \W_\sigma(S)$ then $[\hHam, T]^\sigma \in \W_\sigma(S)$.

\paragraph{Remark:}
Note that according to our definition \eqref{DefAdjoint-2} the wedge condition can also be written, more succinctly, as $\left(\ad^{n+1}\hHam(T)\right)_{-2}=0$. 
This suggests to identify  the graded vector space $\V(S)$ with the subspace of $V_{-2}\oplus \V(S)$ such that all degree $-2$ elements vanishes, i.e. such that $T_{-2}=0$. 
This is trivial addition but it turns out to be very convenient once endowed with an algebra structure. 
It simply requires to extend the $\sigma$-bracket to the degree $-2$ taking $[\cdot\,,\cdot]^\sigma_{-2}:=0$. 
This choice is consistent with the definition of the space $\W_\sigma(S)$ and  perfectly natural since this is precisely what one obtains by evaluating our original definition \eqref{CFbracket} at $s=-2$. 
Notice how on one hand, $[T,T]^\V_{-2}=0$ since the sum is empty, while on the other hand $\paren{T,T'}_{s}=0$ for the special case $s=-2$. \\

The definition of $\hHam$ and its repeated action $\ad^{n+1}\hHam (T)= \big[\hHam,\ad^n\hHam(T)\big]^\sigma$ implies that the action of $\Dcal$ is recursively given by 
\be 
(\Dcal^{n+1} T)_{s}:=D (\Dcal^{n}T)_{s+1}-(s+3)\sigma(\Dcal^{n}T)_{s+2}\,\, \in\,\,\Ccel{(n,-s)},
\ee 
and is such that $\Ham_0^n(\Dcal^n T)_s =\big(\ad^n\hHam(T)\big)_s$.
From there, one straightforwardly obtains that $\big(\ad^{s+2}\hHam (T)\big)_{-2}$ is given to first order in $\sigma$ by the RHS of \eqref{wedge approximate}.
When $\sigma =0$, this condition reduces to the wedge condition $D^{s+2} T_s=0$.
Moreover, since the condition on $T$ is linear, $\W_\sigma(S)$ is a linear space. 
We wrote down already the expression for $(\Dcal^n T)_{-2},\,n=2$, cf. \eqref{doubleB}. 
It is interesting to look at $n=1,3$ too.
In brief, $(\Dcal^n T)_{-2}=0$  for $n=1,2,3$ gives
\bs
\begin{align}
D T_{-1}&=\sigma T_0, \label{Initialcond}\\
D^2\T0 & =2D(\sigma\T1)+\sigma D\T1-3\sigma^2\T2, \\
D^{3} T_1 &= 3 D^2(\sigma T_2)+ 2D(\sigma D\T2) + \sigma D^2 T_2 \cr
& - 8 D(\sigma^2 T_3)- 4 \sigma D(\sigma \T3)- 3 \sigma^2 D T_3 +15 \sigma^3 T_4.
\end{align}
\es

\paragraph{Remark:} 
$\hHam\in\W_\sigma(S)$ because of its non zero central part, namely $\Ham_{-1}$, which has been defined precisely such that $D\Ham_{-1}=D^2G=\sigma=\sigma\Ham_0$ in our preferred frame, and because $\ad\hHam(\hHam)=0$.\\

The previous analysis leads to one of the main results of this paper. 

\begin{tcolorbox}[colback=beige, colframe=argile] \label{theoremWalgebra}
\textbf{Theorem [$\Wcal_\sigma$-algebra]}\\
The space $\Wcal_\sigma(S)\equiv\big(\W_\sigma(S),[\cdot\,,\cdot]^\sigma)$ equipped with the $\sigma$-bracket is a Lie algebra.  
\end{tcolorbox}

\paragraph{Proof:}
The fact that the $\sigma$-bracket satisfies Jacobi follows from \eqref{JacCF} and \eqref{doubleB}.
So far we have also seen from \eqref{algebraRepinterm}, that when  $T,T'\in \W_\sigma$ then $\hat{\delta}_{[T,T']^\sigma}\sigma=0$.
This is a necessary but not sufficient condition for $[T,T']^\sigma$ to belong to $\W_\sigma$.  
To show that the $\sigma$-bracket closes on $\W_\sigma$, i.e. $[T,T']^\sigma\in \W_\sigma$ when $T,T'\in  \W_\sigma$, we first use that when $T,T'\in  \W_\sigma$, then (see App.\,\ref{AppCovWed1})\footnote{Notice that \eqref{doubleB} is compatible with \eqref{com1}.}
\be \label{com1}
\big[\hHam, [T,T']^\sigma\big]^\sigma_{-2}=D[T,T']^\sigma_{-1}- \sigma[T,T']^\sigma_0 
= \Tp0 \hdT\sigma - \T0 \hdTp\sigma=0.
\ee
This in turn allows us to compute the quantity
\begin{align}
    \Big(\ad^2\hHam\big([T,T']^\sigma\big) \Big)_{-2} &=\Big(\ad\hHam\Big(\big[[\hHam,T]^\sigma,T'\big]^\sigma+\big[T,[\hHam,T']^\sigma\big]^\sigma\Big) \Big)_{-2} \cr
    &=\Big(\Tp0\,\hat\delta_{ [\hHam,T]^\sigma} \sigma-[\hHam,T]^\sigma_0\hdTp\sigma\Big)-T\leftrightarrow T' \label{adjointsquare}
    =0.
\end{align}
In the first step, we used that $\hdT\sigma=0=\hdTp\sigma$, i.e. $T,T'\in\Wcal_\sigma(S)$, so that we could leverage the Jacobi identity\footnote{Obviously $\hat\delta_{\hHam}\sigma=0$.} on the  internal elements $\left[\hHam,[T,T']^\sigma\right]^\sigma \cyc 0 $.
The second line follows from \eqref{com1} applied to $(T,T')\to \big([\hHam, T]^\sigma, T'\big)$, while in the last step, we  used again that $T,T'\in\Wcal_\sigma(S)$ together with \eqref{doubleB}.
We can then show recursively that $\big(\ad^n\hHam\big([T,T']^\sigma\big) \big)_{-2}=0$, $n\geqslant 1$, when $T,T'\in\Wcal_\sigma(S)$, which concludes the proof.

\subsection{Covariant form of the $\sigma$-bracket \label{sec:CovFormBracket}}

The fact that $\Wcal_\sigma$ is a Lie algebra suggests that it should be possible to write the bracket in a strict Lie algebra form, namely with field independent structure constants.  
This is what we describe in this subsection, using the  surprising appearance and importance of the degree $-2$ seen in the previous subsection.

Previously, we proved that for $[\cdot\,,\cdot]^\sigma$ to be a Lie bracket, $D\T{-1}$ has to equate $\sigma\T0$. 
Using this condition, we find that 
the $\sigma$-bracket can be simply written as 
\begin{equation}
    \boxed{[T,T']^\sigma_s =\sum_{n=0}^{s+2}(n+1)\big(\T{n}D\Tp{s+1-n}-\Tp{n}D\T{s+1-n}\big)}, \qquad s\geqslant -2. \label{CFbracketSphereVersion}
\end{equation}
The RHS of this expression is similar to the $\V$-bracket, but the sum now runs till $s+2$ and the bracket is naturally defined for all elements $\T{s},~s\geqslant -2$.
The boundary term $n= s+2$ precisely generates the piece $-(s+3)\sigma\paren{T,T'}_s$ once we use the initial constraint $DT_{-1}=\sigma T_0$. Under this constraint we also have that  $[T,T']^\sigma_{-2}= \T0 D\Tp{-1} - \Tp0 DT_{-1}=0$.
It is also worth emphasizing that the degree $-1$ elements are central, as they should. 
Indeed, denoting $\hat c=(c,0,0,\ldots)\in \iota(\V_{-1})\cap\W_\sigma(S)$ (which means that $Dc=0$), we readily get that $[T,\hat c]^\sigma_s=(s+3)\T{s+2}Dc=0$.

\paragraph{Remark:}
This new way of writing $[\cdot\,,\cdot]^\sigma$ sheds light on the fact that $\Wcal_\sigma$ is a Lie \textit{algebra}.
In other words, that \eqref{CFbracketSphereVersion} is a genuine Lie algebra bracket on the sphere, with field \textit{independent} structure constants.
Indeed, so far we said that $\sigma=\sigma(z,\bz)$ was a special field such that $\hdT\sigma=0$. 
Hence, rather than being a structure constant, it played the role of a \textit{functional} structure constant.\footnote{This suggests an algebroid framework to tackle the general case $\hdT\sigma\neq 0$.}
What we show by \eqref{CFbracketSphereVersion} is that the residual field dependency of the functional structure constants can be re-absorbed into the constraint between the generators $\T{-1}$ and $\T0$. 

\paragraph{Remark:}
In the appendix \ref{AppJacobi-1Version}, we give a proof of the Jacobi identity starting from the definition \eqref{CFbracketSphereVersion}, where we stay agnostic about the element $\T{-1}$, i.e. we do not impose $D\T{-1}=\sigma\T0$ a priori.
To avoid confusion, we remove the superscript $\sigma$ to denote the RHS of  \eqref{CFbracketSphereVersion}.
We find that
\begin{align}
    \big[T,[T',T'']\big]_s \cyc - \big[T',T''\big]_{-2} D\T{s+3}-(s+4)\T{s+3}D\big[T',T''\big]_{-2}, \qquad s\geqslant -2, \label{Jacobi-1Version}
\end{align}
which vanishes when $\big[T,T'\big]_{-2}\equiv 0$.
Since 
\begin{equation}
    \big[T,T'\big]_{-2}=\T0 D\Tp{-1}-\Tp0 D\T{-1},
\end{equation}
this happens if and only if the initial constraint \eqref{Initialcond} is satisfied.\footnote{The vanishing of $[T,T']_{-2}$ implies that $DT_{-1}$ is proportional to $T_0$, with a coefficient in $\Ccel{(1,2)}$. This coefficient is $\sigma$.}
Under this condition we have that $[\cdot\,,\cdot]\equiv[\cdot\,,\cdot]^\sigma$, and the analysis of subsec.\,\ref{sec:NewFieldDepBracket} carries over.
This result is interesting on its own since it shows how $[\cdot\,,\cdot]$ could also be a legitimate modification of the $\W$-bracket and another starting point for the analysis performed in subsec.\,\ref{sec:NewFieldDepBracket}.\\

Next, the bracket \eqref{CFbracketSphereVersion} can also be recast as
\begin{equation}
    \boxed{[T,T']^\sigma_s =\sum_{n=0}^{s+1}(n+1)\big(\T{n}(\Dcal T')_{s-n}-\Tp{n}(\Dcal T)_{s-n}\big)}, \qquad s\geqslant -2.\label{CFbracketCovariantSphere}
\end{equation}
To see this, note that the quantity
\begin{equation}
    \sum_{n=0}^{s+2}(n+1)(s+3-n)\sigma\big(\T{n}\Tp{s+2-n}-\Tp{n}\T{s+2-n}\big)
\end{equation}
is equal to its opposite (after a change of variable $n \to s+2-n$), and thus vanishes.
Adding it to \eqref{CFbracketSphereVersion}, it precisely changes $D\to\Dcal$, so that 
\begin{equation}
    [T,T']^\sigma_s =\sum_{n=0}^{s+2}(n+1)\big(\T{n}(\Dcal T')_{s-n}-\Tp{n}(\Dcal T)_{s-n}\big), \qquad s\geqslant -2.\nn
\end{equation}
The term $n\equiv s+2$ of this sum drops since it involves $(\Dcal T)_{-2}$ (and $(\Dcal T')_{-2}$), namely the initial condition \eqref{Initialcond} written in covariant form.

The viewpoint \eqref{CFbracketCovariantSphere} is particularly pleasing since it eventually shows that the $\sigma$-bracket on the covariant wedge space $\W_\sigma(S)$ really is the covariantized version of the $\W$-bracket on the wedge space $\W(S)$. 

Finally, we show that $\Dcal$ is indeed a \textit{derivative operator}, namely that it satisfies the Leibniz rule.

\begin{tcolorbox}[colback=beige, colframe=argile]
\textbf{Lemma [Leibniz rule]} \label{LemmaLeibniz}\\
$\Dcal$ is a covariant derivative and satisfies the Leibniz rule on the covariant wedge algebra bracket $[\cdot\,,\cdot]^\sigma$:
\begin{subequations}
\begin{align}
    & i)\qquad \Dcal\big(f T \big)=f(\Dcal T)+ T Df, \\
    & ii) \!\!\!\qquad \Dcal[T,T']^\sigma 
    =\big[\Dcal T,T'\big]^\sigma +\big[T,\Dcal T'\big]^\sigma, 
\end{align}
\end{subequations}
for $f\in\Ccel{(0,0)}$ a smooth function on $S$.\footnote{To evaluate the first identity at spin $s$ we use that $(T Df)_s= T_{s+1} Df$.}${}^,$\footnote{Strictly speaking, by $ii)$ we mean $\Ham_0\Dcal[T,T']^\sigma=\big[\Ham_0\Dcal T,T'\big]^\sigma+\big[T,\Ham_0\Dcal T'\big]^\sigma$.}
\end{tcolorbox}

\paragraph{Proof:}
$i)$ follows directly from the definition \eqref{CovDer} while $ii)$ follows from the Jacobi identity (that holds when $T,T'\in\Wcal_\sigma(S)$), which ensures that $\ad \hHam =\Ham_0\Dcal$ is a derivation.

\paragraph{Remark:}
For the sake of completeness and as a follow-up to the former remark, we give another demonstration of the property $ii)$ of the previous lemma in appendix \ref{AppLeibniz}.
In this version of the proof, we use the form \eqref{CFbracketCovariantSphere} of the $\sigma$-bracket.
The fact that the covariant wedge condition, now simply written with $\Dcal$ rather than $D$, is preserved by the bracket becomes trivial thanks to the Leibniz rule. 
Indeed,
\begin{equation}
    \big(\Dcal^{s+2}\big[T,T'\big]^\sigma \big)_{-2}=\sum_{k=0}^{s+2}\binom{s+2}{k}\big[\Dcal^k T,\Dcal^{s+2-k} T'\big]^\sigma_{-2}=0
\end{equation}
since the bracket at degree $-2$ is 0 by definition of \eqref{CFbracketCovariantSphere}.

\paragraph{Remark:}
Coming back to the Jacobi identity anomaly \eqref{JacCF} that motivated the whole analysis of the last two subsections, we notice that the most general solution of $\big[T,[T',T'']^\sigma\big]_s^\sigma\cyc 0$ is not $\hat\delta_T\sigma=0$ but 
\be \label{sigmanu}
\hat\delta_T \sigma =-\nu T_0,
\ee
where one has to take $\nu\in\Ccel{(2,2)}$ to get $\nu\T0\in\Ccel{(1,2)}$.
This is due to the fact that  the combination $\nu\T0\paren{T',T''}_s$ vanishes under cyclic permutation, namely
\begin{equation}
    \nu\T0\paren{T',T''}_s = \nu(s+3) \left( \T0\Tp0\Tpp{s+2}-\T0\Tpp0\Tp{s+2}\right)\cyc 0,
\end{equation}
for any $\nu$.
We denote by $\W_{\sigma,\nu}$ the set of elements of $\V(S)$ that satisfy \eqref{sigmanu}.\footnote{More precisely, $\W_{\sigma,\nu}:=\big\{T\in\V(S)~\big|~ (\Dcal T)_{-2}=0~\&~(\Dcal^{s+2}T)_{-2}= \nu(\Dcal^s T)_0,\,s\geqslant 0\big\}$.}
Remarkably, we have that if $T,T'\in \W_{\sigma,\nu}$ then
$[T,T']^\sigma\in \W_\sigma.$ 
This follows from the fact that identities \eqref{com1} and \eqref{adjointsquare} are still satisfied for $T,T' \in \W_{\sigma,\nu}$.
Therefore  the proof given there is still valid and we conclude that $\big(\Dcal^n\big([T,T']^\sigma\big)\big)_{-2} =0$ if $T,T'\in \W_{\sigma,\nu}$, for all $n\geq 0$.
This shows that $\Wcal_\sigma$ is the maximal set which forms a Lie algebra in the sense that $\Wcal_\sigma=[\Wcal_{\sigma,\nu},\Wcal_{\sigma,\nu}]^\sigma$ is the commutator sub-algebra of $\Wcal_{\sigma,\nu}$.

\subsection{Covariant wedge solution \label{sec:CovWedgeSol}}

In this section we show that the two notions of wedge algebra that we have constructed are in fact equivalent: they are related by a Lie algebra isomorphism. 
In practice this requires constructing an element of the covariant wedge  $ \Wcal_\sigma$  given an element of $\Wcal$ and vice-versa. 
This construct relies on the  Goldstone mode $G\in \V_{0}$ introduced in \eqref{Goldstone}.
We also make use of the filtration of $\W_\sigma$
\begin{equation}
    \{0\} \subset\W_\sigma^{-1}\subset \W_{\sigma}^0\subset\W_{\sigma}^1\subset \ldots\subset\W_{\sigma }^s\subset\ldots\subset \W_{\sigma} \quad\textrm{such that}\quad \W_{\sigma}=\bigcup_{n+1\in\mathds{N}}\W_{\sigma }^n,
\end{equation}
where $\W_{\sigma}^s = \W_{\sigma} \cap \V^s$ are the subspaces  of $\W_{\sigma}$ for which $\T{n} = 0$ when $n>s$.\footnote{Note that $\{0\}$ can equivalently be viewed as the trivial space $\W_\sigma^{-2}:= \W_\sigma\cap\V_{-2}$.}
Importantly, the $\sigma$-bracket is of degree $-1$ for this filtration
\be 
[\W_\sigma^s, \W_{\sigma}^{s'}]^\sigma=\W_\sigma^{s+s'-1}.
\ee

Let us start with the definition of the map $\mTG: \hV(S) \to \hV(S) $---where $\hV(S)=\V_{-2}\oplus\V(S)$---given by
\be 
  \boxed{\,\mTG(T):= \sum_{n=0}^{\infty} \frac{G^n}{n!} T^G\big(\ad^n \hHam(T)\big),\,} \label{mTGdef} 
\ee
where $T^G$ is the following filtration preserving map,\footnote{We use the notation $\mTG_s(T)\equiv\big(\mathcal{T}^{G}(T)\big)_s$ and similarly for $T^G$.} 
\be 
T_{s}^G(T):=
\sum_{k=0}^{\infty} 
\frac{(s+k+1)!}{k!(s+1)!} (-DG)^k T_{s+k}, \qquad s\geq -1 \label{TGdef}
\ee 
and is such that $T_{-2}^G(T):=\T{-2}$.\footnote{This can be seen as an extension of the last formula since 
\be
T_{s}^G(T):= T_s + (s+2)
\sum_{k=1}^{\infty} 
\frac{(s+k+1)!}{k!(s+2)!} (-DG)^k T_{s+k},
\ee 
from which we get that $T_{-2}^G(T):= T_{-2}$.} 
Note that if $T\in \hV^p =\bigoplus_{n=-2}^p\V_n$, then both series truncate to finite sums where $n+k\leq p-s$ since  $(\Dcal^nT)_{s+k}=0$ when $s+k+n > p$. In this case we have   $\mTG_p(T) = T_p$ and  the next two terms read
\bs
\begin{align}
\mTG_{p-1}(T) &=
T_{p-1} + G (\Dcal T)_{p-1} - (p+1) DG T_{p} , \\
\mTG_{p-2}(T) &=
T_{p-2} + G (\Dcal T)_{p-2} - p  DG T_{p-1} \cr
& + \frac{G^2}{2} (\Dcal^2 T)_{p-2} - p\, G DG (\Dcal T)_{p-1} + \frac{p(p+1)}{2}
(DG)^2 T_{p}.
\end{align}
\es

This map is linear on $\hV(S)$ and possesses three remarkable properties.

\begin{tcolorbox}[colback=beige, colframe=argile]
\textbf{Lemma [Intertwining properties]} \label{LemmaTG}\\
The map $\mTG$ preserves the filtration, i.e. $\mTG: \hV^s \to \hV^s$; it intertwines the regular and covariant derivative operators
\vspace{-0.1cm}
\begin{subequations}
\begin{align}\label{Dintertwin}
D\mTG(T)& = \mTG\big([\hHam,T]^\sigma\big); 
\end{align}
and  is equal when restricted to $\W_\sigma$ to  the path ordered exponential of the  adjoint action of the element $\hG= (0, G, 0, 0, \cdots) \in \V^0$ along the path $G_t :=e^t G$. 
In other words we have   
\vspace{-0.1cm}
\begin{align}\label{adjointaction}
    \mTG(T) = \overrightarrow{\mathrm{Pexp}}\left(\int_{-\infty}^{0}   \adtG \, \rd t \right)(T),
\end{align}
where  $\sigma_t:=e^t \sigma$ and for $T\in \W_\sigma$.
\end{subequations}
\end{tcolorbox}

\ni The identity \eqref{Dintertwin}  means that $D\mTG_{s+1}(T) = \mTG_{s}\big([\hHam,T]^\sigma\big)$ for $s\geq -2$.
Applying this intertwining equality recursively, we get that 
\be 
D^{s+2} \mTG_s(T) = \mTG_{-2}( \Dcal^{s+2}T) =\sum_{n=0}^{\infty} \frac{G^n}{n!} \big(\Dcal^{s+2+n}T\big)_{-2}\overset{\W_\sigma}{=}0.
\ee
This shows that $\mTG(T) \in \Wcal$ when $T\in \Wcal_\sigma$.

To understand the nature of \eqref{adjointaction}, we recall that the path ordered exponential 
$
W_\alpha:= \overrightarrow{\mathrm{Pexp}}\left(\int_{-\infty}^{\alpha}   A_t \rd t \right)
$ is, by definition, the unique solution to the differential equation $\pa_\alpha W_\alpha = W_\alpha A_\alpha $ with initial condition $W_{-\infty}= 1$. 
From \eqref{adjointaction} we have that
\be \label{PexpT}
\mTaG =  \overrightarrow{\mathrm{Pexp}}\left(\int_{-\infty}^{\alpha}   \adtG \,  \rd t \right),\qquad \Rightarrow \qquad   \pa_{\alpha} \mTaG = \mTaG \circ \adaG.
\ee 
The expression \eqref{adjointaction} also shows that $\mTG$ is invertible with inverse  the opposite path ordered exponential $(\mTG)^{-1} =  \overleftarrow{\mathrm{Pexp}}\left(-\int_{-\infty}^{0}   \adtG \,  \rd t \right)$.
This inverse can be written explicitly as
\be \label{mTinv}
\boxed{\,(\mTG)^{-1}(T)
= 
\sum_{n=0}^{\infty}\frac{(-G)^n}{n!} T^{-G}(D^n T),\,}
\ee
where $T^{-G}$ is the map \eqref{TGdef} with $G\to-G$.
We check in the appendix \ref{app:IInt} that this expression satisfies the intertwining property
$\Dcal (\mTG)^{-1}(T)= (\mTG)^{-1}(DT)$.

The identity \eqref{adjointaction} for $\mTG$ also means that we have the following theorem.

\begin{tcolorbox}[colback=beige, colframe=argile]
\textbf{Theorem [Wedge isomorphism]}\\
$\mTG: \Wcal_\sigma \to \Wcal$,  provides a Lie algebra  isomorphism:
\vspace{-0.1cm}
\begin{align}\label{morphismP}
    \mTG \big([T,T']^\sigma\big)= \left[\mTG(T),\mTG(T')\right]^\W. 
\end{align}
\end{tcolorbox}

\ni 
The morphism property naturally follows from  \eqref{adjointaction}. It is surprising at first since we know that 
$e^{\ad T}$ is an \emph{internal} automorphism of $\Wcal_\sigma$ when $T\in \Wcal_\sigma$.\footnote{In other words $ 
e^{\ad T}\big([T',T'']^\sigma\big) = \big[e^{\ad T}(T'),e^{\ad T}(T'')\big]^{\sigma}$ when $T\in \Wcal_\sigma$.} 
The key point here is that $\hG\notin \W_\sigma $ since it satisfies $\hat{\delta}_{\hG} \sigma = -D^2G =-\sigma$. 
This means that $\ad \hG$ is not a morphism of the $\sigma$-bracket. 
Instead, using the violation of Jacobi identity \eqref{JacCF} on $(G,T,T')$, where $T,T'\in \W_\sigma$, we have that 
\be \label{adG1}
\ad \hat{G} \big([T,T']^\sigma\big) -  \sigma\paren{T,T'}
= \big[\ad \hat{G}(T), T'\big]^\sigma + \big[T,\ad \hG(T')\big]^\sigma. 
\ee 
This equality can be promoted to a differential identity. To do so we  define  
\be T_\alpha := \overleftarrow{\mathrm{Pexp}}\left(-\int_{-\infty}^{\alpha}   \adtG \,  \rd t \right)(T),
\ee 
where $T\in \W$ and similarly for  $T'_\alpha$. By construction 
$T_\alpha, T'_\alpha \in \W_{\sigma_\alpha}$. They are solution of the differential equation 
\be \label{Tader}
\pa_\alpha T_\alpha = - \adaG(T_\alpha),
\ee 
and similarly for $T'_{\alpha}$.
We are now ready to evaluate the following derivative
\begin{align}
& \,\pa_\alpha  
\left( 
\mTaG \left(\left[ T_\alpha, T'_{\alpha}\right]^{\sigma_\alpha}  \right)
\right) \cr
= &\,    
\mTaG \left(\adaG \left(\left[ T_\alpha, T'_{\alpha}\right]^{\sigma_\alpha}  \right)
+ \pa_\alpha \left[ T_\alpha, T'_{\alpha}\right]^{\sigma_\alpha}  
\right) \cr
= &\, 
\mTaG \left(\adaG \left(\left[ T_\alpha, T'_{\alpha}\right]^{\sigma_\alpha}  \right)
- \sigma_\alpha \paren{T_\alpha,T'_{\alpha}} + 
\left[ \pa_\alpha T_\alpha, T'_{\alpha}\right]^{\sigma_\alpha}
+ \left[  T_\alpha, \pa_\alpha T'_{\alpha}\right]^{\sigma_\alpha}
\right) \\
= &\, 
\mTaG \left(\adaG \left(\left[ T_\alpha, T'_{\alpha}\right]^{\sigma_\alpha}  \right)
- \sigma_\alpha \paren{T_\alpha,T'_{\alpha}} - 
\left[ \adaG(T_\alpha), T'_{\alpha}\right]^{\sigma_\alpha}
- \left[  T_\alpha, \adaG(T'_{\alpha})\right]^{\sigma_\alpha}
\right) \cr
=&\, 0. \nn
\end{align}
In the first equality we use \eqref{PexpT}, in the second equality  we use that $\pa_\sigma [T,T']^\sigma = - \paren{T,T'}$, in the third we use the evolution equation  \eqref{Tader} and in the last we leverage \eqref{adG1}.
This shows that $\mTaG \left(\left[ T_\alpha, T'_{\alpha}\right]^{\sigma_\alpha}  \right)$ is independent of $\alpha$. We can evaluate it at $\alpha=0$ and $\alpha=-\infty$. Using that $[\cdot\,,\cdot]^{\sigma=0}\equiv [\cdot\,,\cdot]^{\W}$ and $T_{-\infty}:=T \in  \W$ together with $\mTG(T_{\alpha=0})=T$, we get the morphism property \eqref{morphismP} of $\mTG$.

\paragraph{Proof of $\mTG$ being the path ordered exponential of the adjoint action \eqref{adjointaction}:}
Let us first evaluate the adjoint action of 
$\hat{G}=(0, G, 0,0,\cdots)$ on an arbitrary element $T\in \V(S)$. It is given by  
\vspace{-0.2cm}
\begin{align}
\big(\ad \hat{G} (T)\big)_s=[\hG, T]^\sigma_s = G (\Dcal T)_s - (s+2) DG T_{s+1} \label{ad1G}.
\end{align}
Then we use again the auxiliary parameter $\alpha$ and evaluate in appendix \ref{app:Adj}  the derivative of $\mathcal{T}^{ G_\alpha}$. We use the fact that the $\alpha$  dependence of $\mathcal{T}^{G_\alpha }$  appears directly from the polynomial  dependence
on $G$ and $DG$ but also through the $\sigma$ dependence of $\Dcal\equiv \Dcal_{\sigma} $ since $\sigma= D^2G$.
In the following we denote by $T_\alpha \in \W_{\sigma_\alpha}$ a family of elements which depend smoothly on $\alpha$. We do not assume that $T_\alpha$ satisfies the identity \eqref{Tader}.
We find that
\begin{align}
    \pa_\alpha \big(\mathcal{T}^{ G_\alpha}_s (T_\alpha)\big)
&= \sum_{n,k =0}^{\infty} 
\frac{(s+k+1)!}{n! (s+1)! k!} G_\alpha^{n}(-DG_\alpha)^k
\left(\big(\pa_\alpha + \adaG \big)\big(\Dcal_{\sigma_\alpha}^n T_\alpha \big)  
\right)_{s+k}, \label{palT}
\end{align}
where we used \eqref{ad1G}. 
Next we recast \eqref{adG1} as a differential identity:
\vspace{-0.4cm}
\begin{adjustwidth}{-0.3cm}{0cm}
\be \label{adG2}
\big(\pa_\alpha+ \adaG \big) \big([T_\alpha,T'_\alpha]^{\sigma_\alpha} \big) 
= \big[(\pa_\alpha+ \adaG)(T_\alpha), T'_\alpha \big]^{\sigma_\alpha} + \big[T_\alpha,(\pa_\alpha+ \adaG)(T'_\alpha)\big]^{\sigma_\alpha}. 
\ee 
\end{adjustwidth}
We can rewrite this equality 
in terms of the commutator of the adjoint action as 
\be \label{adG3}
\left[\big(\pa_\alpha + \adaG\big), \ada T_\alpha \right] 
= \ada\big(\pa_\alpha(T_\alpha)+ [ \hat{G}_\alpha,T_\alpha]^{\sigma_\alpha} \big),
\ee 
which is valid when $T_\alpha\in\W_{\sigma_\alpha}$. On the LHS the commutator is simply the commutator of linear operators acting on $\V(S)$.

We can apply this identity for $T_\alpha=\hHama$ and need to remember that  since $\hG_\alpha,\hHama\in \V^0$, their commutator is in $\V_{-1}$ and is therefore a central element. 
Similarly $\pa_\alpha\hHama \in \V_{-1}$ and is central. This implies that
$\big[(\pa_\alpha+ \adaG), \ada \hHama \big] =0$. 
Therefore,
\begin{align}
\big(\pa_\alpha + \adaG \big)\big(\Dcal_{\sigma_\alpha}^n T_\alpha\big) 
& = \big(\pa_\alpha+ \adaG \big)
\big( \ada^n \hHama ( T_\alpha )\big)=  
\ada^n \hHama \big(
(\pa_\alpha + \adaG)(T_\alpha)\big) \cr
& =\Dcal_{ \sigma_\alpha}^n \big(\pa_\alpha  T_\alpha + [\hG_\alpha, T_\alpha]^{\sigma_\alpha}\big). \nn
\end{align} 
From this and \eqref{palT} we conclude that 
$\pa_\alpha \big(\mTaG (T_\alpha)\big) = 
 \mTaG \big( \pa_\alpha T_\alpha +[\hG_\alpha,T_\alpha]^{\sigma_\alpha}\big)$, hence 
\be \label{paltad}
 \pa_\alpha \mTaG  = 
 \mTaG  \circ\adaG.
\ee 
This equation supplemented by the initial condition $\mathcal{T}^0=\mathrm{Id}$ implies that
\be \mTaG=\overrightarrow{\mathrm{Pexp}}\left(\int_{-\infty}^{\alpha}   \adtG \,  \rd t \right).
\ee

\paragraph{Proof of the intertwining property \eqref{Dintertwin}:}
We first show in appendix \ref{app:Int} that the map $T^G$ satisfies the key relationship\footnote{The degree $s-1$ component of this equation reads
$DT_s^G(T) + DG T_{s}^G\big([\hHam, T]^\sigma\big)= T_{s-1}^G\big([\hHam, T]^\sigma\big)$ since $(DG T^G)_{s-1}= DG T_{s}^G$ and $(D T)_{s-1}= D T_s$.}
\be \label{DTrecursion}
DT^G(T) + DG T^G\big([\hHam,T]^\sigma\big)= T^G\big([\hHam, T]^\sigma\big).
\ee 
This in turn implies that 
\begin{align}
 D \mTG(T) 
 &= \sum_{n=0}^{\infty} \frac{G^n}{n!}\Big( D T^G\big(\ad^n \hHam(T)\big)+ DG T^G\big(\ad^{n+1} \hHam(T)\big)\Big) \cr
 &=\sum_{n=0}^{\infty} \frac{G^n}{n!} T^G\big(\ad^{n+1}\hHam(T)\big) = \mTG\big([\hHam, T]^\sigma\big).
\end{align}

\section*{Acknowledgments}
\addcontentsline{toc}{section}{\protect\numberline{}Acknowledgments}

The authors would like to thank Lionel Mason, Romain Ruzziconi and Adam Kmec for many  discussions while they were visiting the Perimeter Institute for the Celestial Summer School 2024. 
We thank Simon Pekar for pointing out a reference on chiral Poincaré.
NC wants to thank Hank Chen for countless hours of discussion and the numerous comments all along the development of this project.
NC would also like to thank Christopher Andrew James Pollack, Adri\'an Lopez and Cole Coughlin for always being open to chat.

\section*{Funding and Competing interests}

The authors declare they have no financial or conflict of interests.
Research at Perimeter Institute is supported by the Government of Canada through the Department of Innovation, Science and Economic Development and by the Province of Ontario through the Ministry of Colleges and Universities. This work was supported by the Simons Collaboration on Celestial Holography.

\section*{Data Availability Statement}

No datasets were generated or analyzed during the current study.

\newpage
\appendix

\section{Solution of the wedge condition on $S_0$ \label{AppWedgeOnSphere}}

We study the solution of the constraint $D^{s+2}\T{s}=0$ on the sphere $S_0$.
Expanding $\T{s}$ in spin-spherical harmonics \eqref{sshExpansion}, we find
\begin{equation}
    \T{-1}=\sum_{\l\geqslant 1}\sum_m T_{-1}^{\l,m}\, Y^1_{\l,m}\qquad\Rightarrow\qquad D\T{-1}=\sum_{\l\geqslant 2} \sqrt{\frac{(\l-1)(\l+2)}{2}}\sum_m T_{-1}^{\l,m}\, Y^2_{\l,m},
\end{equation}
for some coefficients $T_{-1}^{\l,m}\in\C$.
Hence, 
\begin{equation}
    D\T{-1}=0\qquad\Rightarrow\qquad \T{-1}=\sum_{m=-1}^1 T_{-1}^{1,m}\,Y^1_{1,m}. \label{solT-1}
\end{equation}
Similarly,
\begin{equation}
    \T0=\sum_{\l\geqslant 0}\sum_m T_{0}^{\l,m}\, Y^0_{\l,m}\qquad\Rightarrow\qquad D^2\T0=\sum_{\l\geqslant 2}\frac{\sqrt{(\l+2)!}}{2\sqrt{(\l-2)!}}\sum_m T_{0}^{\l,m}\, Y^2_{\l,m}.
\end{equation}
Therefore, 
\begin{equation}
    D^2\T0=0\qquad\Rightarrow\qquad \T0=T^{0,0}_0\,Y^0_{0,0}+\sum_{m=-1}^1 T_0^{1,m}\,Y^0_{1,m}. \label{solT0}
\end{equation}
Note that $Y^0_{0,0}\propto 1$ while  the basis of functions 
$Y^{0}_{1,m}$ is equivalent to the normalized sphere vector $n^i$, with $n^2=1$. Therefore this means that 
\be 
T_0= x^0 + x^i n_i= x^a q_a,
\ee 
for $q_a$ defined by \eqref{T0T}.
In the same vein, we find that
\begin{equation}
    \T1=\sum_{m=-1}^1 T_1^{1,m}Y^{-1}_{1,m}. \label{solT1}
\end{equation}
Then we use \eqref{raisingD} and \eqref{loweringbD} to show that 
\begin{align}
    D Y^{0}_{1,m}  =Y^{1}_{1,m},\qquad 
    \bD Y^{0}_{1,m} =- Y^{-1}_{1,m}.
\end{align}
This establishes \eqref{TDT}.

For $s\geqslant 2$,  one uses that 
\begin{equation}
    \T{s}=\sum_{\l\geqslant s\geqslant 2}\sum_m T_s^{\l,m}\, Y^{-s}_{\l,m}\quad\Rightarrow\quad D^{s+2}\T{s}=\sum_{\l\geqslant s\geqslant 2}\sqrt{\frac{(\l+s)!(\l+2)!}{2^{s+2}(\l-s)!(\l-2)!}}\sum_m T_s^{\l,m}\, Y^2_{\l,m},
\end{equation}
Therefore 
\begin{equation}
    D^{s+2}\T{s}=0\qquad\Rightarrow\qquad \T{s}=0\qquad \textrm{for } s\geqslant 2. \label{solT2}
\end{equation}

\section{Proof of Lemma [Wedge algebra]}

\subsection{Closure \label{AppProofWedge}}

The closure of the $\W$-bracket \eqref{Wbracket} follows from the Leibniz rule which implies that 
\be
D^{s+2} (\T{n} D \Tp{s+1-n})
&=\sum_{k=0}^{s+2}\binom{s+2}{k} D^k \T{n} D^{s+3-k} \Tp{s+1-n} \nn\\
&= \binom{s+2}{n+1}D^{n+1}\T{n}D^{s+2-n}\Tp{s+1-n},
\ee 
To get the second equality, we used that $D^k \T{n}$ vanishes when $k\geq  n+2$ while $D^{s+3-k} \Tp{s+1-n}$ vanishes when $s+3-k \geq s+3 -n$, i.e. when $k\leq n$.
Therefore,
\begin{align}
    D^{s+2}[T,T']^\W_s &= \sum_{n=0}^{s+1}\frac{(s+2)!}{n!(s+1-n)!}\left( D^{n+1}\T{n} D^{s+2-n} \Tp{s+1-n}-D^{n+1}\Tp{n} D^{s+2-n}\T{s+1-n}\right) \nn\\
    &=0, \label{checkWedgeCond}
\end{align}
since by changing $s+1-n\to n$, we see that the sum is equal to its opposite.
This proof holds for $s\geqslant -1$.
This bracket thus acts in the wedge. 

\subsection{Jacobi identity \label{AppJacobiVbracket}}

Here we show that the $\W$-bracket satisfies the Jacobi identity when the wedge condition holds.
\begin{align}
    \big[T,[T',T'']^\W\big]_s^\W &= \sum_{n=0}^{s+1}(n+1)\Big(\T{n}D[T',T'']^\W_{s+1-n}-[T',T'']^\W_n D\T{s+1-n}\Big) \\
    &=\sum_{n=0}^{s+1}\sum_{k=0}^{s+2-n}(n+1)(k+1)\Big(\T{n}D\Tp{k}D\Tpp{s+2-n-k}-\T{n}D\Tpp{k}D\Tp{s+2-n-k} \nn\\
    &\hspace{5cm}+\T{n}\Tp{k}D^2\Tpp{s+2-n-k}-\T{n}\Tpp{k}D^2\Tp{s+2-n-k}\Big) \nn\\
    &-\sum_{n=0}^{s+1}\sum_{k=0}^{n+1}(n+1)(k+1)\Big(\Tp{k}D\Tpp{n+1-k}-\Tpp{k}D\Tp{n+1-k}\Big)D\T{s+1-n}. \nn
\end{align}
It turns out to be very convenient to rewrite the sums as follows
\begin{align}
  \big[T,[T',T'']^\W\big]_s^\W 
    &=\sum_{a+b+c=s+2 \atop b+c>0} \Big( (a+1)(b+1)\T{a}D\Tp{b}D\Tpp{c}-
    (a+1)(c+1)\T{a}D\Tpp{c}D\Tp{b} \Big)\nn\\
    & - \sum_{a+b+c=s+2 \atop b+c>0} \Big( (b+c)(b+1)\Tp{b}D\Tpp{c}D\T{a}-
    (b+c)(c+1)\Tpp{c}D\Tp{b}D\T{a} \Big)\cr
    &+\sum_{a+b+c=s+2 \atop b+c>0}
    \big( (a+1)(b+1)\T{a}\Tp{b}D^2\Tpp{c} -
    (a+1)(c+1)\T{a}\Tpp{c}D^2\Tp{b}\Big).
\end{align} 
In the first two lines we relax the restriction $b+c>0$ as the terms $b=c=0$ and $a =s+2$ vanish. 
In the last line the restriction $b+c>0$ can be relaxed if we  subtract the term\footnote{The latter vanishes inside the $\W$-space. 
For later convenience, we keep it explicit.}
\be 
(s+3)\T{s+2} \big(\Tp0 D^2\Tpp0 - \Tpp0 D^2\Tp0\big).
\ee 
When the summations are simply over $a+b+c= s+2$ we can freely exchange indices. For instance the last line becomes $\sum_{a+b+c=s+2}
     (a+1)(b+1)\big(  \T{a}\Tp{b}D^2\Tpp{c} -\T{a}\Tpp{b}D^2\Tp{c}\Big)$ and therefore  after cyclic permutation we have that 
\begin{align}
  \big[T,[T',T'']^\W\big]_s^\W
    &\cyc\sum_{a+b+c=s+2} \Big( (a+1)(b+1)-
    (a+1)(c+1) \Big) \T{a}D\Tp{b}D\Tpp{c} \nn\\
    & - \sum_{a+b+c=s+2 } \Big( (a+b)(a+1)-
    (a+c)(a+1) \Big) \T{a}D\Tp{b}D\Tpp{c} \nn\\ 
    &+ (s+3) D^2\T0\big(\Tp0 \Tpp{s+2} - \Tp{s+2} \Tpp0\big).
\end{align} 
The first two lines are easily seen to cancel and we are left with \eqref{Jac1}:
\be
\big[T,[T',T'']^\W\big]_s^\W \cyc (s+3) D^2\T0\big(\Tp0 \Tpp{s+2} - \Tp{s+2} \Tpp0\big).
\ee

\section{Poincaré algebra \label{app:Poinc}} 

In this section we establish the results needed for the proof of \eqref{PoincB} and \eqref{nids}. 
One starts with the expression for the unit vector and its derivative 
$Dn = P \pa_z n$, with $P=(1+|z|^2)/\sqrt{2}$:
\be
n&=\frac1{1+|z|^2} \big(z + \bz , -i(z-\bz), 1-|z|^2\big), \cr
Dn&=\frac1{\sqrt{2}} \frac1{1+|z|^2} \big(1- \bz^2 ,-i(1+ \bz^2), -2\bz\big).
\ee 
We then evaluate (notice that $\bD n=\overline{(Dn)}$)
\be
2 D n_1 \bD n_1 + n_1n_1 &= \frac{1}{(1+|z|^2)^2}\left(  (1- \bz^2)(1- z^2) + (z+\bz)^2 \right) = 1,\cr
2 D n_1 \bD n_2 + n_1n_2 &= 
\frac{i}{(1+|z|^2)^2}\left( (1- \bz^2)(1+ z^2) - (z+\bz)(z-\bz) \right) \cr
& = \frac{i}{(1+|z|^2)^2}\left( 1 -|z|^4  \right)  = i n_3.
\ee 
We can evaluate similarly the other components and get
\be 
2 D n_i \bD n_j + n_in_j&= \delta_{ij} + i \epsilon_{ij}{}^k n_k. \label{PoinAlgInterm}
\ee
Besides $D^2 n = \pa_z (P D n)=0$.
From \eqref{casimirDD} we know that $D\bD Y^0_{\ell,m}= -\frac{\ell}{2}(\ell+1) Y^0_{\ell,m}$. Specifying for $\ell=1$ gives $D\bD n_i=-n_i$.
Taking the $D$ derivative of \eqref{PoinAlgInterm}, we thus infer that 
\be 
 n_iDn_j - n_j Dn_i&=  i \epsilon_{ij}{}^k D n_k,
\ee
which establishes \eqref{nids}.

We can now straightforwardly compute the different commutators.
We start with the rotation generators, for which\footnote{Since the notation in terms of the Poincaré generators is clear, we denote for shortness $[\bJ_i,\bJ_j]\equiv \big[\iota(\bJ_i),\iota(\bJ_j)\big]^\W_1$, $[P_a,P_b]\equiv \big[\iota(P_a),\iota(P_b)\big]^\W_{-1}$ and $[\bJ_i,P_b]\equiv \big[\iota(\bJ_i),\iota(P_b)\big]^\W_0$.}
\be 
[\bJ_i,\bJ_j] &= 
2 \big(\bD n_i D\bD n_j - \bD n_j D\bD n_i\big)\cr
&= 2 (n_i \bD n_j - n_j \bD n_i)\cr
&= -2i \epsilon_{ij}{}^k \bD n_k = -2i \epsilon_{ij}{}^k \bJ_k.
\ee 
The translation generators bracket is given by
\be 
[P_a, P_b] &= 
q_a D q_b - q_b D q_a,\cr 
[P_0, P_j] &= 
 q_0 D q_j - q_j D q_0=  D n_j =  C_j,\cr
[P_i, P_j] &= 
q_i D q_j - q_j D q_i= n_iD n_j- n_j Dn_i=
i \epsilon_{ij}{}^k Dn_k
= i \epsilon_{ij}{}^k C_k,
\ee 
where 
we used that $q_0=1$ and $q_i=n_i$.
For the mixed commutators we get 
\be 
[\bJ_i, P_b] &= 
2\bD n_i D q_b - q_b D \bD n_i=  2\bD n_i D q_b + n_i q_b, \cr 
[\bJ_i, P_0 ] &=  n_i = P_i, \cr
[\bJ_i, P_j] &= 
2\bD n_i D n_j + n_i n_j = \delta_{ij} + i \epsilon_{ji}{}^k n_k
= \delta_{ij} P_0 - i\epsilon_{ij}{}^k P_k.
\ee
Using the definition of the 't Hooft symbol \eqref{tHooftSymbol}, we recover \eqref{PoincB}.

\section{Proofs related to the Jacobi identity \label{AppJacobi}}

\subsection{SN bracket \label{JacobiSN}}

As well summarized by \cite{SNbracket}, the fact that the SN bracket satisfies the Jacobi identity is obvious once we know two things. 
Firstly, (and keeping the notation of the \hyperref[SNreminder]{reminder}), that each $S_n\in\X^n(M)$ is uniquely related to a function $f$ on $T^*M$ by 
\begin{equation}
    f(x,p):=S_n(\underbrace{p, \ldots,p}_{n\textrm{ times}})\big|_x = S_n^{k_1\cdots k_n}\big|_x\, p_{k_1} \cdots p_{k_n}\label{HPOdef}
\end{equation}
for all $x\in M$ and co-vector $p\in T^*_x(M)$. 
Such a function is called a \textit{homogeneous polynomial observable of degree n} on $T^*M$.
Given such a function, it determines $S$ uniquely, i.e. \eqref{HPOdef} can be read from right to left too. 
We dub $S_n(f)$ the tensor resulting from this mapping.
Secondly, that the SN bracket \eqref{SNbracketDef} (and by extension \eqref{SNbracketDefbis}) between $S_n(f)$ and $S_q(g)$, for $f$ and $g$ homogeneous polynomial observables of degree $n$ and $q$ respectively, can be defined as 
\begin{equation}
    [S_n(f),S_q(g)]^\sn_{n+q-1}:=S_{n+q-1}\big(\{f,g\}\big),
\end{equation}
where $\{f,g\}=\sum_{i=1}^d\frac{\pa f}{\pa p_i}\frac{\pa g}{\pa x^i}-\frac{\pa g}{\pa p_i}\frac{\pa f}{\pa x^i}$ is the canonical Poisson bracket on $T^*M$ such that $\{p_i,x^j\}=\delta^j_i$.
Jacobi for $[\cdot\,,\cdot]^\sn$ thus follows from Jacobi for $\{\cdot\,,\cdot\}$.

\subsection{Proofs of formulae \eqref{Jacobi1} \& \eqref{Jacobi2} \label{AppJacobiCF}}

We first deal with the two terms proportional to $\sigma$:
\begin{align}
    -\big[T,\sigma\paren{T',T''}\big]_s^\V &- \sigma\paren{T,[T',T'']^\V}_s = \nn\\
    &-\sum_{n=0}^{s+1}(n+1)(s+4-n)\T{n}D\Big(\sigma\big(\Tp0\Tpp{s+3-n}-\Tpp0\Tp{s+3-n}\big)\Big) \nn\\
    & +\sigma\sum_{n=0}^{s+1}(n+1)(n+3)\big(\Tp0\Tpp{n+2}-\Tpp0\Tp{n+2}\big)D\T{s+1-n} \nn\\
    &-(s+3)\sigma\T0\left(\sum_{n=0}^{s+3}(n+1)\big(\Tp{n}D\Tpp{s+3-n}-\Tpp{n}D\Tp{s+3-n}\big)\right) \label{intermJacobi}\\
    &+(s+3)\sigma\T{s+2}\left(\sum_{n=0}^1(n+1) \big(\Tp{n}D\Tpp{1-n}-\Tpp{n}D\Tp{1-n}\big)\right). \nn
\end{align}
In the following, we refer to each line as \circled 1 to \circled 4 respectively. 
First of all,\footnote{Since in this first step we extract the $n=0$ and $n=1$ terms, a whole part of \eqref{eq204} drops if we consider the special cases $s=-1$ and $s=0$.
The final result of the demonstration is however unchanged.}
\begin{align}
    \circled 1 &= -\sum_{n=2}^{s+1}(n+1)(s+4-n)D\sigma\T{n}\big(\Tp0\Tpp{s+3-n}-\Tpp0\Tp{s+3-n}\big) \nn\\
    &\quad -\sigma\sum_{n=2}^{s+1}(n+1)(s+4-n)\T{n}D\big(\Tp0\Tpp{s+3-n}-\Tpp0\Tp{s+3-n}\big) \nn\\
    &\quad -(s+4)\T0 D\big(\sigma(\Tp0\Tpp{s+3}-\Tpp0\Tp{s+3})\big)-2(s+3)\T1 D\big(\sigma(\Tp0\Tpp{s+2}-\Tpp0\Tp{s+2})\big) \nn\\
    &\cyc -\sum_{n=2}^{s+1}(n+1)(s+4-n)D\sigma\Tpp0\big(\cancel{\Tp{n}\T{s+3-n}}-\cancel{\T{n}\Tp{s+3-n}}\big) \label{eq204}\\
    &\quad -\sigma\sum_{n=2}^{s+1}(n+1)(s+4-n)D\T0 \big(\cancel{\Tpp{n}\Tp{s+3-n}}-\cancel{\Tp{n}\Tpp{s+3-n}}\big) \nn\\
    &\quad -\sigma\sum_{n=2}^{s+1}(n+1)(s+4-n)\T0 \big(\Tpp{n}D\Tp{s+3-n}-\Tp{n}D\Tpp{s+3-n}\big) \nn\\
    &\quad +\sigma(s+4)D\T0 \big(\Tp0\Tpp{s+3}-\Tpp0\Tp{s+3}\big)-2(s+3)\T1 D\sigma\big(\Tp0\Tpp{s+2}-\Tpp0\Tp{s+2}\big) \nn\\
    &\quad -2\sigma(s+3)\big(\Tp1 D\Tpp0\T{s+2}-\Tpp1 D\Tp0\T{s+2}\big)-2\sigma(s+3)\big(\Tp1\Tpp0 D\T{s+2}-\Tpp1\Tp0 D\T{s+2}\big), \nn
\end{align}
where we used that
\begin{equation}
    D\big(\sigma\T0\Tp0\Tpp{s+3}-\sigma\T0\Tpp0\Tp{s+3}\big)\cyc 0
\end{equation}
and changed few prime indices for later convenience. 
The sums cancel since equal to their opposite (change variable $n\to s+3-n$ to see it).
Second of all,

\begin{align}
    \circled 2 &\cyc \sum_{n=0}^{s-1}(n+1)(n+3)\sigma\T0\big(\Tp{n+2}D\Tpp{s+1-n}-\Tpp{n+2}D\Tp{s+1-n}\big) \\*
    &+(s+2)(s+4)\sigma D\T0\big(\Tp0\Tpp{s+3}-\Tpp0\Tp{s+3}\big)+ (s+1)(s+3)\sigma D\T1\big(\Tp0\Tpp{s+2}-\Tpp0\Tp{s+2}\big). \nn
\end{align}
Third of all,
\begin{align}
    \circled 3 &\cyc -(s+3)\sigma\T0\left(\sum_{n=2}^{s+1}(n+1)\big(\Tp{n}D\Tpp{s+3-n}-\Tpp{n}D\Tp{s+3-n}\big)\right) \\
    &\quad -(s+3)(s+4)\sigma\big(\Tp0\Tpp{s+3}D\T0-\Tpp0\Tp{s+3}D\T0\big)-(s+3)^2\sigma\big(\Tp0\Tpp{s+2}D\T1-\Tpp0\Tp{s+2}D\T1\big) \nn\\
    &\quad -(s+3)\sigma\big(\cancel{\T0\Tp0 D\Tpp{s+3}}-\cancel{\Tp0\T0 D\Tpp{s+3}}\big)-2(s+3)\sigma\big(\Tp0\Tpp1 D\T{s+2}-\Tpp0\Tp1 D\T{s+2}\big) \nn
\end{align}
Fourth of all,
\begin{equation}
    \circled 4\cyc(s+3)\sigma\big(\Tp{s+2}\Tpp0 D\T1-\Tpp{s+2}\Tp0 D\T1+2\T{s+2}\Tp1 D\Tpp0 -2\T{s+2}\Tpp1 D\Tp0\big).
\end{equation}
Summing all of them,
\begin{align}
    \eqref{intermJacobi} &\cyc\left(-(s+3)+(s+1)(s+3)-(s+3)^2\right)\sigma D\T1\big(\Tp0\Tpp{s+2}-\Tpp0\Tp{s+2}\big) \nn\\
    &+2\big((s+3)-(s+3)\big)\sigma\T{s+2}\big(\Tp1 D\Tpp0 -\Tpp1 D\Tp0\big) \nn\\
    &+\left(-(s+3)(s+4)+(s+4)+(s+2)(s+4)\right) \sigma D\T0\big(\Tp0\Tpp{s+3}-\Tpp0\Tp{s+3}\big) \\
    &+2\big((s+3)-(s+3)\big)\sigma D\T{s+2}\big(\Tp0\Tpp1 -\Tpp0\Tp1\big)-2(s+3)\T1 D\sigma\big(\Tp0\Tpp{s+2}-\Tpp0\Tp{s+2}\big) \nn\\
    &-\sigma\T0\left(\sum_{n=2}^{s+1}\big((n+1)(s+4-n)-(s+3)(n+1)\big)\big(\Tpp{n}D\Tp{s+3-n}-\Tp{n}D\Tpp{s+3-n}\big)\right) \nn\\
    &+\sigma\T0\left(\sum_{n=0}^{s-1}(n+1)(n+3)\big(\Tp{n+2}D\Tpp{s+1-n}-\Tpp{n+2}D\Tp{s+1-n}\big)\right) \nn\\
    &\cyc -3(s+3)\sigma D\T1\big(\Tp0\Tpp{s+2}-\Tpp0\Tp{s+2}\big)-2(s+3)\T1 D\sigma\big(\Tp0\Tpp{s+2}-\Tpp0\Tp{s+2}\big) \nn\\
    &=-\big(3\sigma D\T1+2D\sigma\T1\big)\paren{T',T''}_s. \nn
\end{align}

Concerning the term proportional to $\sigma^2$, we get straightforwardly
\begin{align}
    \sigma \paren{T, \sigma\paren{T',T''}}_s &=(s+3)\sigma^2\left((s+5)\big(\T0\Tp0\Tpp{s+4}-\T0\Tpp0\Tp{s+4}\big)-3(\Tp0\Tpp2-\Tpp0\Tp2)\T{s+2}\right) \nn\\
    &\cyc 3(s+3)\sigma^2\T2\big(\Tp0\Tpp{s+2}-\Tpp0\Tp{s+2}\big) \\
    &\cyc 3 \sigma^2\T2\paren{T',T''}_s. \nn
\end{align}

\section{Proof of the morphism formula \eqref{algebraRepinterm}\label{AppAlgebraAction}}

For convenience, here is the equation \eqref{dTCv1} again:
\begin{equation}
    \hdT \sigma =-D^2\T0+2D\sigma \T1+3\sigma D\T1-3 \sigma ^2\T2.
\end{equation}
This implies that\footnote{$T$ does not depend on $\sigma$, hence $\dTp \T{s}=0,\,s\geqslant 0$.}
\begin{align}
    \big[\hdTp,\hdT\big]\sigma  &=\bigg\{2\T1 D\left(-D^2\Tp0+2D\sigma \Tp1+3\sigma D\Tp1-3 \sigma ^2\Tp2\right) \nn\\
    &\qquad +3D\T1\left(-D^2\Tp0+2D\sigma \Tp1+3\sigma D\Tp1-3 \sigma ^2\Tp2\right) \\
    &\qquad -6\sigma \T2\left(-D^2\Tp0+2D\sigma \Tp1+3\sigma D\Tp1-3 \sigma ^2\Tp2\right)\bigg\} -T\leftrightarrow T'. \nn
\end{align}
On the other hand we know that
\begin{align}
    \hat\delta_{[T,T']^\sigma}\sigma  &=\bigg\{-D^2\left(\T0D\Tp1+2\T1D\Tp0-3 \sigma \T0\Tp2\right) \nn\\
    &\qquad +2D\sigma \big(\T0D\Tp2+2\T1D\Tp1+3\T2D\Tp0-4\sigma \T0\Tp3\big) \\
    &\qquad +3\sigma D\big(\T0D\Tp2+2\T1D\Tp1+3\T2D\Tp0-4\sigma \T0\Tp3\big) \nn\\
    &\qquad -3 \sigma ^2\big(\T0D\Tp3+2\T1D\Tp2+3\T2D\Tp1+4\T3D\Tp0-5 \sigma \T0\Tp4\big)\bigg\}-T\leftrightarrow T'. \nn
\end{align}
Hence, by expanding every term, we compute
\begin{align}
    &\quad ~\big[\hdTp,\hdT\big]\sigma -\hat\delta_{[T,T']^\sigma}\sigma = \nn\\
    &=\Big\{-\cancel{2}\big(\T1D^3\Tp0-\Tp1D^3\T0\big)-\cancel{3}\big(D\T1D^2\Tp0-D\Tp1D^2\T0\big)+\cancel{1}\big(D^2\T0D\Tp1-D^2\Tp0D\T1\big) \nn\\
    &\qquad+\cancel{2}\big(D\T0D^2\Tp1-D\Tp0D^2\T1\big)+\big(\T0D^3\Tp1-\Tp0D^3\T1\big) +\cancel{2}\big(D^2\T1D\Tp0-D^2\Tp1D\T0\big) \nn\\
    &\qquad +\cancel{4}\big(D\T1D^2\Tp0-D\Tp1D^2\T0\big)+\cancel{2}\big(\T1D^3\Tp0-\Tp1D^3\T0\big)\Big\} \nn\\
    &+\Big\{\cancel{4D\sigma }\big(\T1D\Tp1-\Tp1D\T1\big)+\cancel{6\sigma }\big(\T1D^2\Tp1-\Tp1D^2\T1\big)+6\sigma \big(\T2D^2\Tp0-\Tp2D^2\T0\big) \nn\\
    &\quad +\cancel{6D\sigma }\big(\T1D\Tp1-\Tp1D\T1\big)-\cancel{6D\sigma }\big(\T1D\Tp1-\Tp1D\T1\big)-9\sigma \big(\T2D^2\Tp0-\Tp2D^2\T0\big) \nn\\
    &\quad -3D^2\sigma \big(\T0\Tp2-\Tp0\T2\big)-6D\sigma D\big(\T0\Tp2-\Tp0\T2\big)-3\sigma D^2\big(\T0\Tp2-\Tp0\T2\big) \nn\\
    &\quad -2D\sigma \big(\T0D\Tp2-\Tp0D\T2\big)-\cancel{4D\sigma }\big(\T1D\Tp1-\Tp1D\T1\big)-6D\sigma (\T2D\Tp0-\Tp2D\T0\big) \nn\\
    &\quad +6\sigma \big(D\T0D\Tp2-D\Tp0D\T2\big)-3\sigma \big(\T0D^2\Tp2-\Tp0D^2\T2\big)-\cancel{6\sigma }\big(\T1D^2\Tp1-\Tp1D^2\T1\big)\Big\} \nn\\
    &+2\sigma \Big\{-\cancel{6D\sigma }\big(\T1\Tp2-\Tp1\T2\big)-\cancel{3\sigma }\big(\T1D\Tp2-\Tp1D\T2\big)+\cancel{\frac92 \sigma }\big(\T2D\Tp1-\Tp2D\T1\big) \nn\\
    &\qquad~~ -\cancel{6D\sigma }\big(\T2\Tp1-\Tp2\T1\big)-\cancel{9\sigma }\big(\T2D\Tp1-\Tp2D\T1\big)+4D\sigma \big(\T0\Tp3-\Tp0\T3\big) \nn\\
    &\qquad~~ +6D\sigma \big(\T0\Tp3-\Tp0\T3\big)+6\sigma \big(\T0D\Tp3-\Tp0D\T3\big)-\cancel{6\sigma }\big(\T3D\Tp0-\Tp3D\T0\big) \nn\\
    &\qquad~~+\frac32 \sigma \big(\T0D\Tp3-\Tp0D\T3\big)+\cancel{3\sigma }\big(\T1D\Tp2-\Tp1D\T2\big)+\cancel{\frac92 \sigma }\big(\T2D\Tp1-\Tp2D\T1\big) \nn\\
    &\qquad~~+\cancel{6\sigma }\big(\T3D\Tp0-\Tp3D\T0\big)-\frac{15}{2}\sigma ^2\big(\T0\Tp4-\Tp0\T4\big)\Big\}.\nn
\end{align}
We emphasized the obvious simplifications.
A few more occur between the terms involving one $\sigma $.
We notice that we end up only with terms proportional to $\T0$ (and $\Tp0$).

\ni Therefore,
\begin{align}
    &\quad~\big[\hdTp,\hdT\big]\sigma -\hat\delta_{[T,T']^\sigma}\sigma = \nn\\*
    &=\T0\Big\{D^3\Tp1 \underbrace{-3\sigma D^2\Tp2-3D^2\sigma \Tp2-6D\sigma D\Tp2}_{-3D^2(\sigma \Tp2)}-2D\sigma D\Tp2-3\sigma D^2\Tp2+8\sigma D\sigma \Tp3 \nn\\
    &\qquad +12\sigma D\sigma \Tp3+12\sigma ^2D\Tp3+3 \sigma ^2D\Tp3-15\sigma ^3\Tp4\Big\}-T\leftrightarrow T' \nn\\
    &=\T0\Big\{D^2\big(D\Tp1-3\sigma \Tp2\big)-2D\sigma \big(D\Tp2-4\sigma \Tp3\big) \nn\\
    &\qquad -3\sigma D\big(D\Tp2-4\sigma \Tp3\big)+3 \sigma ^2\big(D\Tp3-5 \sigma \Tp4\big)\Big\} -T\leftrightarrow T'\\
    &=\T0\bigg\{D^2(\Dcal T')_0-2D\sigma (\Dcal T')_1-3\sigma D(\Dcal T')_1+3 \sigma^2(\Dcal T')_2\bigg\} -T\leftrightarrow T'\nn\\
    &=\T0\Big\{ -\hat\delta_{[\hHam,T']^\sigma} \sigma\Big\}-T\leftrightarrow T',\nn
\end{align}
which is indeed \eqref{algebraRepinterm}.\footnote{Again, strictly speaking, $\Ham_0\big(\big[\hdTp,\hdT\big]\sigma -\hat\delta_{[T,T']}\sigma\big)=\Tp0 \hat\delta_{[\hHam,T]^\sigma} \sigma-\T0 \hat\delta_{[\hHam,T']^\sigma} \sigma$.}

\section{Covariant wedge initial condition \label{AppCovWed1}}

To prove the formula \eqref{com1}, notice that 
\begin{align}
    D[T,T']^\sigma_{-1} &=D\big(\T0 D\Tp0-2\sigma\T0\Tp1\big)-T\leftrightarrow T' \nn\\
    &=\big(\T0 D^2\Tp0-2D\sigma\T0\Tp1-2\sigma\Tp1 D\T0-2\sigma\T0 D\Tp1\big) -T\leftrightarrow T' \cr
    &=\Big(T_0\big( D^2\Tp0-2D\sigma\Tp1-3\sigma D\Tp1\big) +2\sigma\T1 D\Tp0+\sigma\T0 D\Tp1\Big) -T\leftrightarrow T',
\end{align}
while
\begin{align}
    \sigma[T,T']^\sigma_0 =\big(\sigma\T0 D\Tp1+2\sigma\T1 D\Tp0-3\sigma^2\T0\Tp2\big)-T\leftrightarrow T'.
\end{align}
Therefore, using the definition \eqref{dTCv1} of the variation $\hdT\sigma$, we see that 
\begin{equation}
    D[T,T']^\sigma_{-1}- \sigma[T,T']^\sigma_0 = \Tp0\hdT\sigma-\T0\hdTp\sigma,
\label{IniConstInterm}
\end{equation}
which matches with \eqref{com1} up to the usual factor of $\Ham_0$ that we keep implicit.

\section{Proof of the Jacobi identity \eqref{Jacobi-1Version} \label{AppJacobi-1Version}}

Here we study the condition that the bracket \eqref{CFbracketSphereVersion} has to satisfy in order to respect the Jacobi identity, if we were not aware that $D\T{-1}$ has to be equal to $\sigma\T0$.
To avoid confusion, we remove the superscript $\sigma$ to denote this preliminary version of the $\sigma$-bracket.
\begin{align}
    \big[T,[T',T'']\big]_s &=\sum_{n=0}^{s+2}(n+1)\Big(\T{n}D[T',T'']_{s+1-n}-[T',T'']_n D\T{s+1-n}\Big) \label{Jac-1interm}\\
    &= \sum_{n=0}^{s+2}\sum_{k=0}^{s+3-n}(n+1)(k+1)\Big(\T{n}D\Tp{k}D\Tpp{s+2-n-k}-\T{n}D\Tpp{k}D\Tp{s+2-n-k} \nn\\
    &\hspace{5cm}+\T{n}\Tp{k}D^2\Tpp{s+2-n-k}-\T{n}\Tpp{k}D^2\Tp{s+2-n-k}\Big) \nn\\
    &-\sum_{n=0}^{s+2}\sum_{k=0}^{n+2}(n+1)(k+1)\Big(\Tp{k}D\Tpp{n+1-k}-\Tpp{k}D\Tp{n+1-k}\Big)D\T{s+1-n}. \nn
\end{align}
The goal is to extract the boundary terms of the sums, namely those that involve the element $\T{-1}$.
Therefore
\begin{align}
    \big[T,[T',T'']\big]_s &=\sum_{a+b+c=s+2}(a+1)(b+1)\Big(\T{a}D\Tp{b}D\Tpp{c}-\T{a}D\Tpp{b}D\Tp{c}\Big) \nn\\
    &+\sum_{n=0}^{s+2}(n+1)(s+4-n)\Big(\T{n}D\Tp{s+3-n}D\Tpp{-1}-\T{n}D\Tpp{s+3-n}D\Tp{-1}\Big) \nn\\
    &+\sum_{a+b+c=s+2}(a+1)(b+1)\Big(\T{a}\Tp{b}D^2\Tpp{c}-\T{a}\Tpp{b}D^2\Tp{c}\Big) \nn\\
    &+\sum_{n=0}^{s+2}(n+1)(s+4-n)\Big(\T{n}\Tp{s+3-n}D^2\Tpp{-1}-\T{n}\Tpp{s+3-n}D^2\Tp{-1}\Big) \nn\\
    &-\sum_{a+b+c=s+2}(b+c)(b+1)\Big(\Tp{b}D\Tpp{c}D\T{a}-\Tpp{b}D\Tp{c}D\T{a}\Big) \\
    &-\sum_{k=0}^{s+4}(s+3)(k+1)\Big(\Tp{k}D\Tpp{s+3-k}D\T{-1}-\Tpp{k}D\Tp{s+3-k}D\T{-1}\Big) \nn\\
    &-\sum_{n=0}^{s+1}(n+1)(n+3)\Big(\Tp{n+2}D\Tpp{-1}D\T{s+1-n}-\Tpp{n+2}D\Tp{-1}D\T{s+1-n}\Big). \nn
\end{align}
According to \eqref{Jac-1interm}, the first sum that comes with a minus sign should be restricted to $b+c>0$, but of course we can include the term $b+c=0$ for free since that contribution vanishes. 
Upon cyclic permutations, we get
\begin{align}
    \big[T,[T',T'']\big]_s &\cyc \!\!\!\!\!\sum_{a+b+c=s+2}\!\!\!\!\!\T{a}D\Tp{b}D\Tpp{c} \big((a+1)(b+1)-(a+1)(c+1)-(a+b)(a+1)+(a+c)(a+1)\big) \nn\\
    &+\sum_{a+b+c=s+2}\T{a}\Tp{b}D^2\Tpp{c}\big( (a+1)(b+1)-(b+1)(a+1)\big) \nn\\
    &+D\T{-1}\left\{\sum_{n=0}^{s+2}(n+1)(s+4-n)\Big(\Tp{n}D\Tpp{s+3-n}-\Tpp{n}D\Tp{s+3-n}\Big)\right. \nn\\
    &\qquad\quad~ -\sum_{n=0}^{s+4}(s+3)(n+1)\Big(\Tp{n}D\Tpp{s+3-n}-\Tpp{n}D\Tp{s+3-n}\Big) \\
    &\qquad\quad~ -\left.\sum_{n=0}^{s+1}(n+1)(n+3)\Big(\Tpp{n+2}D\Tp{s+1-n}-\Tp{n+2}D\Tpp{s+1-n}\Big)\right\} \nn\\
    &-D^2\T{-1}(s+4)\Big(\Tp{s+3}\Tpp0-\Tpp{s+3}\Tp0\Big). \nn
\end{align}
The first two lines vanish. Hence
\begin{align}
    \big[T,[T',T'']\big]_s &\cyc D\T{-1}\left\{\sum_{n=0}^{s+2}(n+1)(1-n)\Big(\Tp{n}D\Tpp{s+3-n}-\Tpp{n}D\Tp{s+3-n}\Big)\right. \nn\\
    & \qquad\quad~ -(s+3)(s+4)\Big(\Tp{s+3}D\Tpp0-\Tpp{s+3}D\Tp0\Big) \\
    &\quad\qquad~ -\left.\sum_{n=0}^{s+1}(n+1)(n+3)\Big(\Tpp{n+2}D\Tp{s+1-n}-\Tp{n+2}D\Tpp{s+1-n}\Big)\right\} \nn\\
    &-D^2\T{-1}(s+4)\Big(\Tp{s+3}\Tpp0-\Tpp{s+3}\Tp0\Big), \nn
\end{align}
where we got rid of
\begin{equation}
    -D\T{-1}(s+3)(s+5)\big(\Tp{s+4}D\Tpp{-1}-\Tpp{s+4}D\Tp{-1}\Big)\cyc 0.
\end{equation}
Massaging the sums (as usual by extracting some boundary terms), we have
\begin{align}
    \big[T,[T',T'']\big]_s &\cyc  D\T{-1}\bigg\{\Tp0 D\Tpp{s+3}-\Tpp0 D\Tp{s+3}-\sum_{n=2}^{s+2}(n+1)(n-1)\Big(\Tp{n}D\Tpp{s+3-n}-\Tpp{n}D\Tp{s+3-n}\Big) \nn\\
    &\qquad\quad~ -\sum_{n=2}^{s+2}(n-1)(n+1)\Big(\Tpp{n}D\Tp{s+3-n}-\Tp{n}D\Tpp{s+3-n}\Big) \\
    & \qquad\quad~ -(s+4)\Big(\Tp{s+3}D\Tpp0-\Tpp{s+3}D\Tp0\Big)\bigg\} -D^2\T{-1}(s+4)\Big(\Tp{s+3}\Tpp0-\Tpp{s+3}\Tp0\Big), \nn
\end{align}
The two sums cancel one another and we can also get rid of the $D^2\T{-1}$ term using the Leibniz's rule (over the last term in $\{\cdots\}$), so that one ends up with
\begin{align}
    \big[T,[T',T'']\big]_s &\cyc -(s+4)D\Big(D\T{-1}\Tpp0\Tp{s+3}-D\T{-1}\Tp0\Tpp{s+3}\Big)+(s+3)D\T{-1}\Big(\Tpp0 D\Tp{s+3}-\Tp0 D\Tpp{s+3}\Big) \nn\\
    &\cyc -(s+4)D\Big(\T{s+3}\big(\Tp0 D\Tpp{-1}-\Tpp0 D\Tp{-1}\big)\Big)+(s+3)D\T{s+3}\Big(\Tp0 D\Tpp{-1}-\Tpp0 D\Tp{-1}\Big) \nn\\
    &\cyc -D\T{s+3}\big[T',T''\big]_{-2}-(s+4)\T{s+3}D\big[T',T''\big]_{-2}.
\end{align}
In the second line we used the cyclic permutation one more time to highlight the combination
\begin{equation}
    \big[T',T''\big]_{-2}=\Tp0 D\Tpp{-1}-\Tpp0 D\Tp{-1}.
\end{equation}
This concludes the proof of \eqref{Jacobi-1Version}.

\section{Proof of Lemma [Leibniz] \label{AppLeibniz}}

We start by computing explicitly $\big(\Dcal[T,T']^\sigma\big)_{s-1}$ using the formula \eqref{CFbracketCovariantSphere} for the $\sigma$-bracket. 
\begin{align}
    \big(\Dcal[T,T']^\sigma\big)_{s-1} &=D[T,T']^\sigma_s-(s+2)\sigma [T,T']_{s+1}^\sigma \nn\\
    &=\left\{\sum_{n=0}^{s+1}(n+1)\Big(D\T{n}(\Dcal T')_{s-n}+\T{n}D(\Dcal T')_{s-n}\Big)\right. \\
    & \left.-(s+2)\sigma\sum_{n=0}^{s+2}(n+1)\T{n}(\Dcal T')_{s+1-n}\right\}-(T\leftrightarrow T'). \nn
\end{align}
We now add and subtract the necessary terms to transform the $D$ into $\Dcal$ in the first sum:
\begin{align}
    \big(\Dcal[T,T']^\sigma\big)_{s-1}  &= \left\{\sum_{n=0}^{s+1}(n+1)\Big((\Dcal T)_{n-1}(\Dcal T')_{s-n}+\T{n}(\Dcal^2 T')_{s-n-1}\Big)\right. \nn\\
    &+\sum_{n=0}^{s+1}(n+1)(n+2)\sigma\T{n+1}(\Dcal T')_{s-n}+\sum_{n=0}^{s+1}(n+1)(s-n+2)\sigma\T{n}(\Dcal T')_{s-n+1} \nn\\
    &-\left.(s+2)\sigma\sum_{n=0}^{s+1}(n+1)\T{n}(\Dcal T')_{s+1-n}-(s+2)(s+3)\sigma\T{s+2}(\Dcal T')_{-1}\right\} -(T\leftrightarrow T') \nn\\
    &=\left\{\sum_{n=0}^{s+1}(n+1)\Big((\Dcal T)_{n-1}(\Dcal T')_{s-n}+\T{n}(\Dcal^2 T')_{s-n-1}\Big)\right. \nn\\
    &\quad +\sum_{n=1}^{s+2}n(n+1)\sigma\T{n}(\Dcal T')_{s+1-n}-\sum_{n=1}^{s+1}n(n+1)\sigma\T{n}(\Dcal T')_{s+1-n} \nn\\
    &\quad -(s+2)(s+3)\sigma\T{s+2}(\Dcal T')_{-1}\Bigg\} -(T\leftrightarrow T') \\
    &=\left\{\sum_{n=0}^{s+1}(n+1)\Big((\Dcal T)_{n-1}(\Dcal T')_{s-n}+\T{n}(\Dcal^2 T')_{s-n-1}\Big)\right\} -(T\leftrightarrow T'). \nn
\end{align}
Extracting the boundary term $n=0$ in the $\Dcal T\Dcal T'$ sum (and renaming $n-1\rightarrow n$), we can write the result as:
\begin{align}
    \big(\Dcal[T,T']^\sigma\big)_{s-1} &=\left\{\sum_{n=0}^{s+1}(n+1)\T{n}(\Dcal^2 T')_{s-n-1} +\sum_{n=0}^s(n+2)(\Dcal T)_n(\Dcal T')_{s-n-1}\right. \nn\\
    &\quad +(\Dcal T)_{-1}(\Dcal T')_s\Bigg\}-(T\leftrightarrow T').
\end{align}
We then split the $\Dcal T\Dcal T'$ sum as $n+2=(n+1)+1$ and use the condition $(\Dcal^2 T')_{-2}=0$ to drop the top term of the first sum.
\begin{align}
    \big(\Dcal[T,T']^\sigma\big)_{s-1} &=\left\{\sum_{n=0}^{s}(n+1)\T{n}(\Dcal^2 T')_{s-n-1} +\sum_{n=0}^s(n+1)(\Dcal T)_n(\Dcal T')_{s-n-1}\right\}-(T\leftrightarrow T') \nn\\
    &+\left\{\sum_{n=0}^s(\Dcal T)_n(\Dcal T')_{s-n-1}+(\Dcal T)_{-1}(\Dcal T')_s\right\}-(T\leftrightarrow T') \\
    &=\left\{\sum_{n=0}^{s}(n+1)\T{n}(\Dcal^2 T')_{s-n-1} -\sum_{n=0}^s(n+1)(\Dcal T')_n(\Dcal T)_{s-n-1}\right\}-(T\leftrightarrow T') \nn\\
    &=\big[T,\Dcal T'\big]^\sigma_{s-1}-(T\leftrightarrow T') \nn\\
    &=\big[T,\Dcal T'\big]^\sigma_{s-1}+\big[\Dcal T,T'\big]^\sigma_{s-1},
\end{align}
where we used the fact that
\begin{align}
    &\quad\left\{\sum_{n=0}^s(\Dcal T)_n(\Dcal T')_{s-n-1}+(\Dcal T)_{-1}(\Dcal T')_s\right\}-(T\leftrightarrow T') \nn\\
    &=\left\{\sum_{n=0}^{s-1}(\Dcal T)_n(\Dcal T')_{s-n-1}\right\}-(T\leftrightarrow T')=0.
\end{align}
The last equality follows after the renaming $n \to s-1-n$. 
This overall demonstration is valid for $s\geqslant 0$. 
Since we also proved that $\big(\Dcal[T,T']^\sigma\big)_{-2}=0$ when $(\Dcal^2 T)_{-2}=0$ in App.\,\ref{AppCovWed1}, this concludes the proof of the \hyperref[LemmaLeibniz]{Lemma}.

\section{Proof of Lemma [Intertwining properties]}
\subsection{Adjoint action}
\label{app:Adj} 

We evaluate the derivative
\begin{align}
    \pa_\alpha \big(\mathcal{T}^{ G_\alpha}_s (T_\alpha)\big)
    &= 
\sum_{n,k =0}^{\infty} 
\frac{(s+k+1)!}{n! (s+1)! k!} G_\alpha^{n}(-DG_\alpha)^k \Big( (n+k)  
\big(\Dcal_{\sigma_\alpha}^nT_\alpha\big)_{s+k} + \pa_\alpha \big(\Dcal_{\sigma_\alpha}^nT_\alpha\big)_{s+k}\Big) \cr
    &= 
\sum_{n=1}^\infty\sum_{k=0}^{\infty} \left(
\frac{(s+k+1)!}{(n-1)! (s+1)! k!} G_\alpha^{n}(-DG_\alpha)^k\right)\big(\Dcal_{\sigma_\alpha}^nT_\alpha\big)_{s+k} \nn\\
&+\sum_{n=0}^\infty\sum_{k=1}^{\infty} \left( 
\frac{(s+k+1)!}{n! (s+1)! (k-1)!} G_\alpha^n(-DG_\alpha)^{k} \right)\big(\Dcal_{\sigma_\alpha}^nT_\alpha\big)_{s+k} \cr
&+ \sum_{n,k =0}^{\infty} 
\frac{(s+k+1)!}{n! (s+1)! k!} G_\alpha^{n}(-DG_\alpha)^k \left( \pa_\alpha \big(\Dcal_{\sigma_\alpha}^nT_\alpha\big)_{s+k}\right)\cr
&= \sum_{n,k =0}^{\infty} 
\frac{(s+k+1)!}{n! (s+1)! k!} G_\alpha^{n}(-DG_\alpha)^k
\Big(\!
\big(G_\alpha \Dcal_{\sigma_\alpha}^{n+1}T_\alpha\big)_{s+k}+\pa_\alpha \big(\Dcal_{\sigma_\alpha}^nT_\alpha \big)_{s+k} \nn\\
&\hspace{6cm}-(s+k+2) DG_\alpha \big(\Dcal^n_{\sigma_\alpha} T_\alpha\big)_{s+k+1}\Big) \nn\\
&= \sum_{n,k =0}^{\infty} 
\frac{(s+k+1)!}{n! (s+1)! k!} G_\alpha^{n}(-DG_\alpha)^k
\left(\big(\pa_\alpha+\adaG\big) \big(\Dcal_{\sigma_\alpha}^n T_\alpha\big)
\right)_{s+k},
\end{align} 
which corresponds to \eqref{palT}.

\subsection{Intertwining relation}
\label{app:Int}
Here we prove \eqref{DTrecursion}.
Starting from 
\be 
T_{s}^G(T):=
\sum_{k=0}^\infty 
\frac{(s+k+1)!}{(s+1)! k!} (-DG)^k T_{s+k},\qquad s\geqslant -1,
\ee 
we easily see that we have for $T\in\V^p$,\footnote{If $T\in \V^p$ then the series truncates to a finite sum where $k\leq p-s$ since  $T_{s+k}=0$ for $ k > p-s$.}
\begin{align} 
DT_{s}^G(T)& =
\sum_{k=0}^{p-s}
\frac{(s+k+1)!}{(s+1)! k!} \Big((-DG)^k D T_{s+k} - k (-DG)^{k-1} D^2G \,T_{s+k}\Big) \cr
& =
\sum_{k=0}^{p-s} 
\frac{(s+k+1)!}{(s+1)! k!} (-DG)^k D T_{s+k}
- \sum_{k=1}^{p-s} 
\frac{(s+k)!}{(s+1)! (k-1)!}  (-DG)^{k-1}  (s+k+1) \sigma  T_{s+k}\cr
& =
\sum_{k=0}^{p-s} 
\frac{(s+k+1)!}{(s+1)! k!} (-DG)^k D T_{s+k}-\sum_{k=0}^{p-s-1} 
\frac{(s+k+1)!}{(s+1)! k!} (-DG)^k  (s+k+2) \sigma  T_{s+k+1} \cr
& =
\sum_{k=0}^{p-s} 
\frac{(s+k+1)!}{(s+1)! k!} (-DG)^k  (\Dcal T)_{s+k-1}.
\end{align}
Next we evaluate ($\Dcal$ shifts the filtration since it comes from a bracket, i.e. $\Dcal:\V^p\to\V^{p-1}$)
\begin{align}
    &\quad\big(D T^G(T)\big)_{s-1} + \big(DG T^G\big([\hHam,T]^\sigma\big)\big)_{s-1} =
    D T_{s}^G(T) + D G T^G_{s}\big([\hHam,T]^\sigma\big) \cr
    &= \sum_{k=0}^{p-s} 
\frac{(s+k+1)!}{(s+1)! k!}(-DG)^k  (\Dcal T)_{s+k-1} - \sum_{k=0}^{p-s-1} 
\frac{(s+k+1)!}{(s+1)! k!} (-DG)^{k+1}  (\Dcal T)_{s+k} \cr
&= (\Dcal T)_{s-1} + \sum_{k=1}^{p-s} 
\left(\frac{(s+k+1)!}{(s+1)! k!}- 
\frac{(s+k)!}{(s+1)! (k-1)!} \right)  (-DG)^k  (\Dcal T)_{s+k-1} \label{diffs}.
\end{align}
At this stage there are two cases to consider: 
If $s=-1$ then we see that the two coefficients appearing in the difference inside the sum are equal to $1$. 
Their contributions cancel and we are left with
\be \label{TG-2interm}
D T_{-1}^G(T) + DG T_{-1}^G\big([\hHam,T]^\sigma\big)= (\Dcal T)_{-2}=T^G_{-2}\big([\hHam, T]^\sigma \big).
\ee
When $s>-1$ we can evaluate the difference and get that \eqref{diffs} reduces to
\be 
(\Dcal T)_{s-1} + \sum_{k=1}^{p-s} 
\frac{(s+k)!}{s! k!}(-DG)^k  (\Dcal T)_{s+k-1}. \label{DTG}
\ee
On the other hand we have that 
\be \label{Teta}
   T_{s-1}^G\big([\hHam, T]^\sigma \big) =
    \sum_{k=0}^{p-s} 
\frac{(s+k)!}{s! k!} (-DG)^k  (\Dcal T)_{s+k-1}.
\ee
Equation \eqref{TG-2interm} and the equality of \eqref{DTG} with \eqref{Teta} proves \eqref{DTrecursion}.

\subsection{Inverse intertwining relation}
\label{app:IInt}

Here we prove  that $(\mTG)^{-1}$ defined in \eqref{mTinv} satisfies the intertwining relation  $\Dcal (\mTG)^{-1}(T) = (\mTG)^{-1}(DT)$.
Using that
\be 
T_{s}^{-G}(T)=
\sum_{k=0}^\infty 
\frac{(s+k+1)!}{(s+1)! k!} (DG)^k T_{s+k},\qquad s\geqslant -1,
\ee
one evaluates 
\begin{align}
    \big(\Dcal T^{-G}(T)\big)_{s-1} &= 
    DT_{s}^{-G}(T) - (s+2) \sigma T_{s+1}^{-G}(T)\cr
    &=\sum_{k=1}^\infty\frac{(s+1+k)!}{(s+1)!(k-1)!}(DG)^{k-1}D^2G\,\T{s+k}+\sum_{k=0}^\infty\frac{(s+1+k)!}{(s+1)!k!}(DG)^k D\T{s+k} \cr
& - \sigma 
\sum_{k=0}^{\infty} 
\frac{(s+k+2)!}{(s+1)! k!}
(DG)^k T_{s+k+1}
\cr
&=\sum_{k=0}^{\infty} 
\frac{(s+k+1)!}{(s+1)! k!} (DG)^k  (D T)_{s+k-1},
\end{align}
where in the second line, the first and third terms cancel out since $D^2G=\sigma$.
Next we evaluate\footnote{For $s=-1$, $\big(\Dcal T^{-G}(T)- D G T^{-G}(DT)\big)_{-2}=(DT)_{-2}=T^{-G}_{-2}(DT)$.}
\begin{align}
    &\quad\big(\Dcal T^{-G}(T)\big)_{s-1} - \big(DG T^{-G}(D T)\big)_{s-1} =
    \big(\Dcal T^{-G}(T)\big)_{s-1} - D G T^{-G}_{s}(DT) \cr
    &= \sum_{k=0}^{\infty} 
\frac{(s+k+1)!}{(s+1)! k!}(DG)^k  (D T)_{s+k-1} - \sum_{k=0}^{\infty} 
\frac{(s+k+1)!}{(s+1)! k!} (DG)^{k+1}  (D T)_{s+k} \cr
&= (DT)_{s-1} + \sum_{k=1}^\infty 
\left(\frac{(s+k+1)!}{(s+1)! k!}- 
\frac{(s+k)!}{(s+1)! (k-1)!} \right)  (DG)^k  (D T)_{s+k-1} \cr
& =
    \sum_{k=0}^\infty 
\frac{(s+k)!}{s! k!} (DG)^k  (D T)_{s+k-1}
=  T_{s-1}^{-G}(DT),\qquad\qquad s\geqslant 0.
\end{align}
This in turn implies that 
\begin{align}
 \Dcal (\mTG)^{-1}(T) 
 &= \sum_{n=0}^{\infty} \frac{(-G)^n}{n!}\Big( \Dcal T^{-G}\big(D^n T \big)- DG T^{-G}\big(D^{n+1} T \big)\Big) \cr
 &=\sum_{n=0}^{\infty} \frac{(-G)^n}{n!} T^{-G}\big(D^{n+1}T\big) = (\mTG)^{-1}(DT).
\end{align}

\newpage
\addcontentsline{toc}{section}{References}

\bibliographystyle{JHEP}
\bibliography{Arxiv_Covariant_Wedge_final_v3}

\providecommand{\href}[2]{#2}\begingroup\raggedright\begin{thebibliography}{10}

\bibitem{Guevara:2021abz}
A.~Guevara, E.~Himwich, M.~Pate and A.~Strominger, \emph{{Holographic symmetry algebras for gauge theory and gravity}}, \href{https://doi.org/10.1007/JHEP11(2021)152}{\emph{JHEP} {\bfseries 11} (2021) 152} [\href{https://arxiv.org/abs/2103.03961}{{\ttfamily 2103.03961}}].

\bibitem{Strominger:2021mtt}
A.~Strominger, \emph{{$w_{1+\infty}$ Algebra and the Celestial Sphere: Infinite Towers of Soft Graviton, Photon, and Gluon Symmetries}}, \href{https://doi.org/10.1103/PhysRevLett.127.221601}{\emph{Phys. Rev. Lett.} {\bfseries 127} (2021) 221601} [\href{https://arxiv.org/abs/2105.14346}{{\ttfamily 2105.14346}}].

\bibitem{Himwich:2021dau}
E.~Himwich, M.~Pate and K.~Singh, \emph{{Celestial operator product expansions and $w_{1+\infty}$ symmetry for all spins}}, \href{https://doi.org/10.1007/JHEP01(2022)080}{\emph{JHEP} {\bfseries 01} (2022) 080} [\href{https://arxiv.org/abs/2108.07763}{{\ttfamily 2108.07763}}].

\bibitem{Ball:2021tmb}
A.~Ball, S.A.~Narayanan, J.~Salzer and A.~Strominger, \emph{{Perturbatively exact $w_{1+\infty}$ asymptotic symmetry of quantum self-dual gravity}}, \href{https://doi.org/10.1007/JHEP01(2022)114}{\emph{JHEP} {\bfseries 01} (2022) 114} [\href{https://arxiv.org/abs/2111.10392}{{\ttfamily 2111.10392}}].

\bibitem{zamolodchikov1995infinite}
A.B.~Zamolodchikov, \emph{Infinite additional symmetries in two-dimensional conformal quantum field theory}, {\emph{W-Symmetry. World Scientific} (1995) 221}.

\bibitem{PseudoDiffBakas}
I.~Bakas, \emph{Higher spin fields and the gelfand-dickey algebra}, \href{https://doi.org/10.1007/bf01218588}{\emph{Communications in Mathematical Physics 1989-dec vol. 123 iss. 4} {\bfseries 123} (1989) }.

\bibitem{BAKAS198957}
I.~Bakas, \emph{The large-n limit of extended conformal symmetries}, \href{https://doi.org/https://doi.org/10.1016/0370-2693(89)90525-X}{\emph{Physics Letters B} {\bfseries 228} (1989) 57}.

\bibitem{Boyer_1985}
C.P.~Boyer and J.F.~Plebański, \emph{{An infinite hierarchy of conservation laws and nonlinear superposition principles for self‐dual Einstein spaces}}, \href{https://doi.org/10.1063/1.526652}{\emph{Journal of Mathematical Physics} {\bfseries 26} (1985) 229} [\href{https://arxiv.org/abs/https://pubs.aip.org/aip/jmp/article-pdf/26/2/229/19107151/229\_1\_online.pdf}{{\ttfamily https://pubs.aip.org/aip/jmp/article-pdf/26/2/229/19107151/229\_1\_online.pdf}}].

\bibitem{Shen:1992dd}
X.~Shen, \emph{{W infinity and string theory}}, \href{https://doi.org/10.1142/S0217751X92003203}{\emph{Int. J. Mod. Phys. A} {\bfseries 7} (1992) 6953} [\href{https://arxiv.org/abs/hep-th/9202072}{{\ttfamily hep-th/9202072}}].

\bibitem{Dunajski:2000iq}
M.~Dunajski and L.J.~Mason, \emph{{HyperKahler hierarchies and their twistor theory}}, \href{https://doi.org/10.1007/PL00005532}{\emph{Commun. Math. Phys.} {\bfseries 213} (2000) 641} [\href{https://arxiv.org/abs/math/0001008}{{\ttfamily math/0001008}}].

\bibitem{Adamo:2021lrv}
T.~Adamo, L.~Mason and A.~Sharma, \emph{{Celestial $w_{1+\infty}$ Symmetries from Twistor Space}}, \href{https://doi.org/10.3842/SIGMA.2022.016}{\emph{SIGMA} {\bfseries 18} (2022) 016} [\href{https://arxiv.org/abs/2110.06066}{{\ttfamily 2110.06066}}].

\bibitem{Mason:2022hly}
L.~Mason, \emph{{Gravity from holomorphic discs and celestial $Lw_{1+\infty }$ symmetries}}, \href{https://doi.org/10.1007/s11005-023-01735-2}{\emph{Lett. Math. Phys.} {\bfseries 113} (2023) 111} [\href{https://arxiv.org/abs/2212.10895}{{\ttfamily 2212.10895}}].

\bibitem{Penrose:1976js}
R.~Penrose, \emph{{Nonlinear gravitons and curved twistor theory}}, \href{https://doi.org/10.1007/BF00762011}{\emph{Gen. Rel. Grav.} {\bfseries 7} (1976) 31}.

\bibitem{Adamo:2021bej}
T.~Adamo, L.~Mason and A.~Sharma, \emph{{Twistor sigma models for quaternionic geometry and graviton scattering}}, \href{https://doi.org/10.4310/ATMP.2023.v27.n3.a1}{\emph{Adv. Theor. Math. Phys.} {\bfseries 27} (2023) 623} [\href{https://arxiv.org/abs/2103.16984}{{\ttfamily 2103.16984}}].

\bibitem{Bu:2022iak}
W.~Bu, S.~Heuveline and D.~Skinner, \emph{{Moyal deformations, W$_{1+\infty}$ and celestial holography}}, \href{https://doi.org/10.1007/JHEP12(2022)011}{\emph{JHEP} {\bfseries 12} (2022) 011} [\href{https://arxiv.org/abs/2208.13750}{{\ttfamily 2208.13750}}].

\bibitem{Mason:2023mti}
L.~Mason, R.~Ruzziconi and A.~Yelleshpur~Srikant, \emph{{Carrollian amplitudes and celestial symmetries}}, \href{https://doi.org/10.1007/JHEP05(2024)012}{\emph{JHEP} {\bfseries 05} (2024) 012} [\href{https://arxiv.org/abs/2312.10138}{{\ttfamily 2312.10138}}].

\bibitem{Taylor:2023ajd}
T.R.~Taylor and B.~Zhu, \emph{{w1+\ensuremath{\infty} Algebra with a Cosmological Constant and the Celestial Sphere}}, \href{https://doi.org/10.1103/PhysRevLett.132.221602}{\emph{Phys. Rev. Lett.} {\bfseries 132} (2024) 221602} [\href{https://arxiv.org/abs/2312.00876}{{\ttfamily 2312.00876}}].

\bibitem{Bittleston:2024rqe}
R.~Bittleston, G.~Bogna, S.~Heuveline, A.~Kmec, L.~Mason and D.~Skinner, \emph{{On AdS$_{4}$ deformations of celestial symmetries}}, \href{https://doi.org/10.1007/JHEP07(2024)010}{\emph{JHEP} {\bfseries 07} (2024) 010} [\href{https://arxiv.org/abs/2403.18011}{{\ttfamily 2403.18011}}].

\bibitem{Bondi:1962px}
H.~Bondi, M.G.J.~van~der Burg and A.W.K.~Metzner, \emph{{Gravitational waves in general relativity. 7. Waves from axisymmetric isolated systems}}, \href{https://doi.org/10.1098/rspa.1962.0161}{\emph{Proc. Roy. Soc. Lond.} {\bfseries A269} (1962) 21}.

\bibitem{Sachs:1962wk}
R.K.~Sachs, \emph{{Gravitational waves in general relativity. 8. Waves in asymptotically flat space-times}}, \href{https://doi.org/10.1098/rspa.1962.0206}{\emph{Proc. Roy. Soc. Lond.} {\bfseries A270} (1962) 103}.

\bibitem{Strominger:2017zoo}
A.~Strominger, \emph{{Lectures on the Infrared Structure of Gravity and Gauge Theory}} (3, 2017), [\href{https://arxiv.org/abs/1703.05448}{{\ttfamily 1703.05448}}].

\bibitem{Raclariu:2021zjz}
A.-M.~Raclariu, \emph{{Lectures on Celestial Holography}},  \href{https://arxiv.org/abs/2107.02075}{{\ttfamily 2107.02075}}.

\bibitem{Hamada:2018vrw}
Y.~Hamada and G.~Shiu, \emph{{Infinite Set of Soft Theorems in Gauge-Gravity Theories as Ward-Takahashi Identities}}, \href{https://doi.org/10.1103/PhysRevLett.120.201601}{\emph{Phys. Rev. Lett.} {\bfseries 120} (2018) 201601} [\href{https://arxiv.org/abs/1801.05528}{{\ttfamily 1801.05528}}].

\bibitem{Freidel:2021dfs}
L.~Freidel, D.~Pranzetti and A.-M.~Raclariu, \emph{{Sub-subleading soft graviton theorem from asymptotic Einstein\textquoteright{}s equations}}, \href{https://doi.org/10.1007/JHEP05(2022)186}{\emph{JHEP} {\bfseries 05} (2022) 186} [\href{https://arxiv.org/abs/2111.15607}{{\ttfamily 2111.15607}}].

\bibitem{Freidel:2021ytz}
L.~Freidel, D.~Pranzetti and A.-M.~Raclariu, \emph{{Higher spin dynamics in gravity and \ensuremath{w_{1+\infty}} celestial symmetries}}, \href{https://doi.org/10.1103/PhysRevD.106.086013}{\emph{Phys. Rev. D} {\bfseries 106} (2022) 086013} [\href{https://arxiv.org/abs/2112.15573}{{\ttfamily 2112.15573}}].

\bibitem{Geiller:2024bgf}
M.~Geiller, \emph{Celestial $w_{1+\infty}$ charges and the subleading structure of asymptotically-flat spacetimes},  \href{https://arxiv.org/abs/2403.05195}{{\ttfamily 2403.05195}}.

\bibitem{Schouten}
J.A.~Schouten, \emph{\"uber differentialkonkomitanten zweier kontravarianten gr\"ossen}, {\emph{Mathematics} {\bfseries 2} (1940) 449}.

\bibitem{SNbracket}
L.K.~Norris, \emph{Schouten-nijenhuis brackets}, \href{https://doi.org/10.1063/1.531981}{\emph{Journal of Mathematical Physics} {\bfseries 38} (1997) 2694} [\href{https://arxiv.org/abs/https://pubs.aip.org/aip/jmp/article-pdf/38/5/2694/19202771/2694\_1\_online.pdf}{{\ttfamily https://pubs.aip.org/aip/jmp/article-pdf/38/5/2694/19202771/2694\_1\_online.pdf}}].

\bibitem{Barnich:2016lyg}
G.~Barnich and C.~Troessaert, \emph{{Finite BMS transformations}}, \href{https://doi.org/10.1007/JHEP03(2016)167}{\emph{JHEP} {\bfseries 03} (2016) 167} [\href{https://arxiv.org/abs/1601.04090}{{\ttfamily 1601.04090}}].

\bibitem{Campiglia:2014yka}
M.~Campiglia and A.~Laddha, \emph{{Asymptotic symmetries and subleading soft graviton theorem}}, \href{https://doi.org/10.1103/PhysRevD.90.124028}{\emph{Phys. Rev. D} {\bfseries 90} (2014) 124028} [\href{https://arxiv.org/abs/1408.2228}{{\ttfamily 1408.2228}}].

\bibitem{Compere:2018ylh}
G.~Comp\`ere, A.~Fiorucci and R.~Ruzziconi, \emph{{Superboost transitions, refraction memory and super-Lorentz charge algebra}}, \href{https://doi.org/10.1007/JHEP11(2018)200}{\emph{JHEP} {\bfseries 11} (2018) 200} [\href{https://arxiv.org/abs/1810.00377}{{\ttfamily 1810.00377}}].

\bibitem{Campiglia:2020qvc}
M.~Campiglia and J.~Peraza, \emph{{Generalized BMS charge algebra}}, \href{https://doi.org/10.1103/PhysRevD.101.104039}{\emph{Phys. Rev. D} {\bfseries 101} (2020) 104039} [\href{https://arxiv.org/abs/2002.06691}{{\ttfamily 2002.06691}}].

\bibitem{Freidel:2021fxf}
L.~Freidel, R.~Oliveri, D.~Pranzetti and S.~Speziale, \emph{{The Weyl BMS group and Einstein\textquoteright{}s equations}}, \href{https://doi.org/10.1007/JHEP07(2021)170}{\emph{JHEP} {\bfseries 07} (2021) 170} [\href{https://arxiv.org/abs/2104.05793}{{\ttfamily 2104.05793}}].

\bibitem{Donnay:2024qwq}
L.~Donnay, L.~Freidel and Y.~Herfray, \emph{{Carrollian \ensuremath{Lw_{1+\infty}} representation from twistor space}},  \href{https://arxiv.org/abs/2402.00688}{{\ttfamily 2402.00688}}.

\bibitem{Krasnov:2021cva}
K.~Krasnov and E.~Skvortsov, \emph{{Flat self-dual gravity}}, \href{https://doi.org/10.1007/JHEP08(2021)082}{\emph{JHEP} {\bfseries 08} (2021) 082} [\href{https://arxiv.org/abs/2106.01397}{{\ttfamily 2106.01397}}].

\bibitem{Freidel:2023gue}
L.~Freidel, D.~Pranzetti and A.-M.~Raclariu, \emph{On infinite symmetry algebras in yang-mills theory}, \href{https://doi.org/10.1007/JHEP12(2023)009}{\emph{JHEP} {\bfseries 12} (2023) 009} [\href{https://arxiv.org/abs/2306.02373}{{\ttfamily 2306.02373}}].

\bibitem{Kmec:2024nmu}
A.~Kmec, L.~Mason, R.~Ruzziconi and A.~Yelleshpur~Srikant, \emph{{Celestial $Lw_{1+\infty}$ charges from a twistor action}},  \href{https://arxiv.org/abs/2407.04028}{{\ttfamily 2407.04028}}.

\bibitem{newman_heaven_1976}
E.~Newman, \emph{Heaven and its properties}, {\emph{General Relativity and Gravitation} {\bfseries 7} (1976) 107}.

\bibitem{Adamo:2009vu}
T.M.~Adamo, C.N.~Kozameh and E.T.~Newman, \emph{{Null Geodesic Congruences, Asymptotically Flat Space-Times and Their Physical Interpretation}}, \href{https://doi.org/10.12942/lrr-2009-6}{\emph{Living Rev. Rel.} {\bfseries 12} (2009) 6} [\href{https://arxiv.org/abs/0906.2155}{{\ttfamily 0906.2155}}].

\bibitem{Cresto:2024mne}
N.~Cresto and L.~Freidel, \emph{{Asymptotic Higher Spin Symmetries II: Noether Realization in Gravity}},  \href{https://arxiv.org/abs/2410.15219}{{\ttfamily 2410.15219}}.

\bibitem{ethop}
J.N.~Goldberg, A.J.~MacFarlane, E.T.~Newman, F.~Rohrlich and E.C.G.~Sudarshan, \emph{{Spin s spherical harmonics and edth}}, \href{https://doi.org/10.1063/1.1705135}{\emph{J. Math. Phys.} {\bfseries 8} (1967) 2155}.

\bibitem{francesco2012conformal}
P.~Francesco, P.~Mathieu and D.~S{\'e}n{\'e}chal, \emph{Conformal field theory}, Springer Science \& Business Media (2012).

\bibitem{Barnich:2021dta}
G.~Barnich and R.~Ruzziconi, \emph{{Coadjoint representation of the BMS group on celestial Riemann surfaces}}, \href{https://doi.org/10.1007/JHEP06(2021)079}{\emph{JHEP} {\bfseries 06} (2021) 079} [\href{https://arxiv.org/abs/2103.11253}{{\ttfamily 2103.11253}}].

\bibitem{Donnay:2021wrk}
L.~Donnay and R.~Ruzziconi, \emph{{BMS flux algebra in celestial holography}}, \href{https://doi.org/10.1007/JHEP11(2021)040}{\emph{JHEP} {\bfseries 11} (2021) 040} [\href{https://arxiv.org/abs/2108.11969}{{\ttfamily 2108.11969}}].

\bibitem{Eastwood_Tod_1982}
M.~Eastwood and P.~Tod, \emph{Edth-a differential operator on the sphere}, \href{https://doi.org/10.1017/S0305004100059971}{\emph{Mathematical Proceedings of the Cambridge Philosophical Society} {\bfseries 92} (1982) 317–330}.

\bibitem{gelfand1966generalized}
I.~Gel'fand, M.~Graev and N.Y.~Vilenkin, \emph{Generalized functions. Vol. 5: Integral geometry and representation theory}, Academic Press New York and London (1966).

\bibitem{t1976computation}
G.~t~Hooft, \emph{Computation of the quantum effects due to a four-dimensional pseudoparticle}, {\emph{Physical Review D} {\bfseries 14} (1976) 3432}.

\bibitem{Nakayama:2023xzu}
Y.~Nakayama, \emph{{Central Extension of Scaling Poincar\'e Algebra}},  \href{https://arxiv.org/abs/2311.10254}{{\ttfamily 2311.10254}}.

\bibitem{moser1965volume}
J.~Moser, \emph{On the volume elements on manifolds}, {\emph{Trans. Amer. Math. Soc.} {\bfseries 120} (1965) 280}.

\end{thebibliography}\endgroup
\end{document}